\documentclass[journal=inoraj,manuscript=article,layout=traditional]{achemso}
\usepackage{chemformula} % Formula subscripts using \ch{} as in my .bib file.
\usepackage{gensymb}
\usepackage{graphicx}
\usepackage{booktabs}
\usepackage{multirow}
\usepackage{longtable}
\usepackage{dcolumn}% Align table columns on decimal point
\usepackage{bm}% bold math
\usepackage{hyperref}% add hypertext capabilities
\usepackage{booktabs}% nice tables
\usepackage{textcomp} 

\usepackage{xcolor} %For colouring some text

\SectionNumbersOn

\title{Pressure tuning the Jahn-Teller transition temperature in NaNiO$_2$}
\author{Liam A. V. Nagle-Cocco}
\email{lavn2@cam.ac.uk}
\affiliation{Cavendish Laboratory, University of Cambridge, JJ Thomson Avenue, Cambridge, CB3 0HE, United Kingdom.}

\author{Craig L. Bull}
\affiliation{ISIS Neutron and Muon Facility, Rutherford Appleton Laboratory, Didcot, OX11 0QX, United Kingdom.}
\affiliation{School of Chemistry, University of Edinburgh, David Brewster Road, Edinburgh, EH9 3FJ, United Kingdom.}

\author{Christopher J. Ridley}
\affiliation{ISIS Neutron and Muon Facility, Rutherford Appleton Laboratory, Didcot, OX11 0QX, United Kingdom.}

\author{Si\^an E. Dutton}
\email{sed33@cam.ac.uk}
\affiliation{Cavendish Laboratory, University of Cambridge, JJ Thomson Avenue, Cambridge, CB3 0HE, United Kingdom.}

\begin{document}
	
	

	\begin{abstract}
		\textbf{Abstract:} NaNiO$_2$ is a layered material consisting of alternating layers of NaO$_6$ and Jahn-Teller-active NiO$_6$ edge-sharing octahedra. At ambient pressure it undergoes a broad phase transition from a monoclinic to rhombohedral structure between $\sim$465\,K and $\sim$495\,K, associated with the loss of long-range orbital ordering. In this work, we present the results of a neutron powder diffraction study on powdered NaNiO$_2$ as a function of pressure and temperature from ambient pressure to $\sim$5\,GPa between 290\,K and 490\,K. 
		The 290\,K and 460\,K isothermal compressions remained in the monoclinic phase up to the maximum pressures studied, whereas the 490\,K isotherm was mixed-phase throughout. 
		The unit-cell volume was fitted to a 2nd-order Birch-Murnaghan equation of state, with $B=113(1)$\,GPa. 
		We observe at 490\,K that the fraction of Jahn-Teller-distorted phase increases with increasing pressure, from 67.8(6)\% at 0.71(2)\,GPa to 80.2(9)\% at 4.20(6)\,GPa. Using this observation, in conjunction with neutron diffraction measurements at 490\,K from 5.46(9)\,GPa to 0.342(13)\,GPa, we show that the Jahn-Teller transition temperature increases with pressure. Our results are used to present a structural pressure-temperature phase diagram for NaNiO$_2$.
		To our knowledge, this is the first diffraction study of the effect of pressure on the Jahn-Teller transition temperature in materials with edge-sharing Jahn-Teller-distorted octahedra, and the first variable-pressure study on a JT-active edge-sharing nickelate.
	\end{abstract}
	
	\maketitle
	
	\section{\label{sec:level1}Introduction}
	
	Many transition metal oxides exhibit a Jahn-Teller (JT) distortion due to degeneracy in the 3$d$ orbitals, manifesting as an elongation or compression of the \textit{M}O$_6$ (\textit{M}=transition metal ion) octahedra, generally with associated orbital ordering. Previous studies on the effect of pressure on materials containing JT-active ions have found that pressure can entirely suppress the JT distortion and orbital ordering~\cite{bhadram2021reentrant,zhou2008breakdown}. It has also been observed that application of pressure reduces the magnitude of distortion in $M$O$_6$ octahedra~\cite{loa2001pressure,zhou2011jahn,caslin2016competing,collings2018disorder,bostrom2019high}.
	
	One well-studied material under pressure is LaMnO$_3$ (with JT active \textit{d}$^4$ Mn$^{3+}$ ions)~\cite{loa2001pressure,pinsard2001stability,zhou2008breakdown}. %An insulator at low pressures, the material exhibits a metal-to-insulator transition at 32\,GPa~\cite{loa2001pressure}. 
	At ambient pressure it adopts the perovskite structure with corner-sharing MnO$_6$ octahedra. An ordered JT distortion results in an orthorhombic symmetry. At $T> \sim 750$\,K the JT distortion is suppressed, and there is an increase in symmetry first to a cubic phase with octahedral tilting, and then at higher temperatures to a rhombohedral phase~\cite{rodriguez1998neutron}. The temperature-driven suppression of the JT distortion coincides with a marked increase in electronic conductivity~\cite{zhou1999paramagnetic}. On application of pressure at room temperature $P < 8$\,GPa, the JT distortion is decreased through reduction of the long Mn-O bond lengths~\cite{loa2001pressure}. At $\sim$11\,GPa a rhombohedral phase with no JT distortion co-exists with the distorted orthorhombic phase~\cite{zhou2008breakdown}, becoming single-phase at $\sim$12\,GPa.
	Similarly, the manganese (III) quadruple perovskite LaMn$_7$O$_{12}$, exhibits a complete suppression of the JT distortion at $\sim$34\,GPa~\cite{bhadram2021reentrant}. %A study on fluoride perovskite KCuF$_3$, which features JT distorted $d^9$ CuF$_6$ octahedra, showed the difference between long and short Cu-F bond lengths decreasing with increasing pressure, and similar to LaMnO$_3$ they do not reach equality, and there is no observed suppression of $T_\mathrm{JT}$ to room temperature, within the measured pressure range ($<8$\,GPa)~\cite{zhou2011jahn}. 
				
	There are several interesting studies on JT-distorted compounds with edge-sharing octahedra. Here we describe three different examples classes, all containing JT-active $d^4$ Mn$^{3+}$. Mn$_3$O$_4$, a spinel containing both Mn$^{3+}$ and Mn$^{2+}$, has been found to exhibit different pressure-dependence of JT-distorted octahedra depending on morphology; for instance, in single-crystal Mn$_3$O$_4$ the JT disortion survives to 60\,GPa~\cite{ovsyannikov2021structural}, whereas there are observed transitions to JT-free phases at much lower pressures in powdered~\cite{paris1992mn3o4} and nanorod~\cite{li2020size} Mn$_3$O$_4$. ZnMnO$_2$, also with spinel-type structure but with Zn$^{2+}$ on the Mn$^{2+}$ of Mn$_3$O$_4$, has been studied to very high pressure ($\sim 52$\,GPa)~\cite{aasbrink1999high}, with a transition reported at $\sim 23$\,GPa which has been alternately described as a transition from JT-elongation to a slight JT-compression~\cite{aasbrink1999high} or a spin-crossover transition resulting in an insulator$\rightarrow$metal transition~\cite{choi2006electronic}. CuMnO$_2$, with delafossite structure, has also had the pressure-dependence of its JT distortion studied~\cite{lawler2021decoupling}. It exhibits a higher compressibility in the long Mn-O bond than the short Mn-O bond similar to LaMnO$_3$~\cite{loa2001pressure} and other materials~\cite{zhou2011jahn,caslin2016competing} up to around around $\sim 10$\,GPa; above this pressure there is an isostructural phase transition associated with a collapse in the interlayer (\textit{c}-axis) and an increase in the volume of the Mn$^{3+}$O$_6$ JT-distorted octahedra.
		
	Nickelates containing JT-active $d^7$ Ni$^{3+}$ are far less studied than the manganates under pressure. This may be partly because many materials containing $d^7$ Ni$^{3+}$ octahedra do not exhibit a co-operative JT distortion, where the JT distortion is long-range ordered. NdNiO$_3$, which has been subject to a variable-pressure structural study~\cite{medarde1997pressure}, is not considered to contain a JT distortion~\cite{garcia1994neutron,mizokawa2000spin,garcia2009structure}, as is the case for most Ni-containing perovskites~\cite{johnston2014charge}. Similarly, AgNiO$_2$ is widely accepted to not contain any kind of JT distortion~\cite{wawrzynska2007orbital,kang2007valence}. LiNiO$_2$ is an interesting case as it does not display long-range magnetic or orbital ordering, likely due to Li/Ni site mixing; some experimental results have been interpreted as evidence for a non-cooperative JT distortion~\cite{rougier1995non,chung2005local}, although this is debated~\cite{chen2011charge,foyevtsova2019linio,green2020evidence}. Similarly various nickel-containing perovskites~\cite{zhou2004chemical} are subject to discussion regarding whether there exists any kind of JT distortion.

	NaNiO$_2$ is a layered $d^7$ nickelate. The presence of the JT distortion in NaNiO$_2$ is not subject to debate~\cite{dick1997structure,chappel2000study,green2020evidence}, even amongst proponents of alternative theories for degeneracy-breaking in LiNiO$_2$~\cite{green2020evidence}. NaNiO$_2$ is therefore an ideal choice for studying the effect of pressure on the JT distortion in a material which is both a nickelate and has edge-sharing octahedra. The room-temperature phase of NaNiO$_2$ is a semiconductor, based on its black colour and by analogy with LiNiO$_2$~\cite{galakhov1995electronic}, but we do not know of any measurement of the conductivity properties of the high-temperature phase. NaNiO$_2$ is of interest because of its magnetic ground state, consisting at ambient pressure of intra-layer ferromagnetism and inter-layer anti-ferromagnetism~\cite{lewis2005ordering,baker2005thermodynamic,darie2005magnetic}. It has also been studied in recent years because $A$NiO$_2$ ($A = \mathrm{alkali\ metal}$) is the template compound for Ni-rich alkali metal-transition metal oxides within the field of batteries~\cite{vassilaras2012electrochemical,han2014structural}.
		
\begin{figure*}[t]
			
		\makebox[\textwidth][c]{\includegraphics[trim=0 0cm 3.3cm 1.9cm,clip,width=220mm]{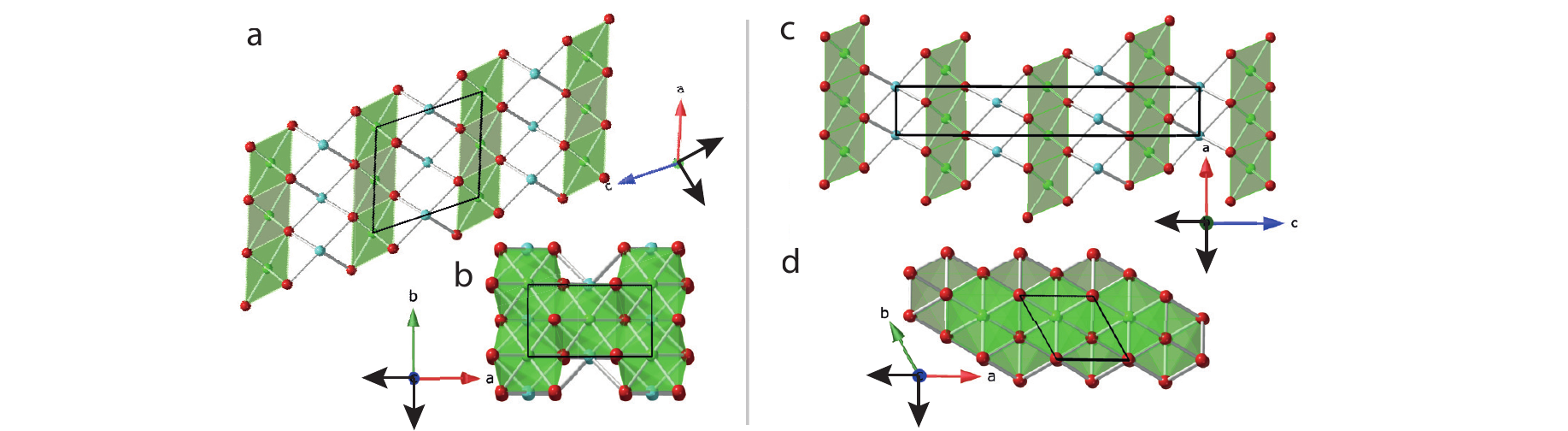}}
		\caption{\label{struc_fig}(a) and (b) show the monoclinic, Jahn-Teller distorted NaNiO$_2$ phase along the $b$-axis and $c$-axis respectively. (c) and (d) show the rhombohedral, JT-inactive NaNiO$_2$ phase along the $b$ and $c$ axes respectively. Ni$^{3+}$ cations are shown in green, O$^{2-}$ anions in red, and Na$^+$ cations in cyan. Na$^+$ ions and octahedra are hidden in (b) and (d). The solid black quadrilaterals denote the unit cell. The black arrows represent the directions of principal axes of compression projected into the (a,c) $ac$-plane and (b,d) $ab$-plane.}

	\end{figure*}

	NaNiO$_2$ has an ordered JT distortion at room temperature due to degeneracy in $e_g$ orbitals in low-spin Ni$^{3+}$. It exhibits a first-order phase transition between 465\,K and 495\,K to an undistorted phase. The crystal structures are shown in Figure~\ref{struc_fig}. The monoclinic ($C2/m$) JT-distorted phase consists of alternating layers of edge-sharing NiO$_6$ and NaO$_6$ octahedra. Both the NiO$_6$ and NaO$_6$ octahedra exhibit angular and bond-length distortions from geometrically regular octahedra. Ni, Na, and O ions occupy the 2\textit{a}(0,0,0), 2\textit{d}(0,$\frac{1}{2}$,$\frac{1}{2}$), and 4\textit{i}($x$,0,$z$) Wyckoff sites respectively. The rhombohedral ($R\bar{3}m$) phase consists of the same arrangement of alternating NiO$_6$ and NaO$_6$ layers of edge-sharing octahedra, with octahedra bound within layers by O$_4$ tetrahedra. In this phase, Ni, Na, and O ions occupy the 3\textit{b}(0,0,$\frac{1}{2}$), 3\textit{a}(0,0,0), and 6\textit{c}(0,0,$z$) Wyckoff sites respectively. The unit cell remains centrosymmetric, with the change in symmetry due solely to the repression of the JT distortion and the resulting non-variable $r_\mathrm{\textit{M}-O}$ bond lengths.
	
	In this work, we present a structural study of NaNiO$_2$ as a function of temperature between 290\,K and 500\,K, and pressure up to 5.46(9)\,GPa. We demonstrate using the 490\,K isotherm that the JT transition temperature increases between 2\,GPa and 4.2\,GPa, increasing more rapidly with pressure at higher pressures, while the degree of distortion decreases over this pressure range.
	
	\section{\label{sec:level2}Methods}
		
	\textit{Sample preparation and characterisation.} Samples were prepared by solid state synthesis. Na$_2$O$_2$ (Alfa Aesar; 95\%) and NiO (Alfa Aesar; 99.995\%) were mixed and pelletised in a 1.05:1 molar ratio of Na:Ni, with excess Na to account for Na-loss during heating. Sample was heated to 973\,K for 70 hrs in a tube furnace under constant flow of O$_2$. To prevent reaction with moisture, the sample was stored and handled in an inert Ar-atmosphere. X-Ray Diffraction (XRD) data were obtained using a Bruker D8 Discover powder (Cu K$\alpha_{1,2}$, $\lambda = 1.541$\,\AA) diffractometer. A Mira3 TESCAN Scanning Electron Microscopy was used to obtain SEM images of the morphology of NaNiO$_2$, with an accelerating electron voltage of 3\,kV (for SEM images, see SI).
	
	\textit{Ambient-pressure neutron diffraction.} Ambient-pressure neutron diffraction was performed using the NOMAD instrument~\cite{neuefeind2012nanoscale} at the Spallation Neutron Source, Oak Ridge National Laboratory, USA. NaNiO$_2$ was sealed in a glass ampoule for the measurements. Heating was performed using a furnace. The sample was measured during heating at 293\,K, 450\,K, 500\,K, and after cooling at 316\,K.
	
	\textit{Variable-pressure neutron diffraction.}	Variable-temperature and -pressure neutron diffraction studies were performed at the PEARL instrument~\cite{bull2016pearl} ($2\theta=90^\circ$), ISIS Neutron and Muon Source, UK, using a V3 Paris-Edinburgh press. The sample was measured between 0.107(8)\,GPa and 4.24(5)\,GPa at 290\,K, 0.130(10)\,GPa and 5.29(8)\,GPa at 460\,K, and 0.254(17)\,GPa and 4.20(6)\,GPa at 490\,K. NaNiO$_2$ was packed into a encapsulated null scattering TiZr gasket which was loaded in a zirconia-toughened alumina toroidal profile anvil, with a lead pellet for pressure calibration~\cite{fortes_manual_paper}. Anhydrous deuterated methanol:ethanol (4:1 by volume) was used as a pressure transmitting medium for the ambient temperature isothermal compression experiment. Preliminary measurements indicated that NaNiO$_2$ reacted with the methanol:ethanol solution at higher temperatures [SI Figure~S2], and so a 1:1 ratio by volume of FC77:FC84 fluorinert (purchased from 3M) was used for the 460\,K and 490\,K isotherms. The data were processed and corrected using	Mantid~\cite{arnold2014mantid}.
		
	\textit{Diffraction Analysis.} Diffraction data were analysed using the software package \textsc{topas 5}~\cite{coelho2018topas}, utilising Pawley fitting~\cite{pawley1981unit} and Rietveld refinement~\cite{rietveld1969profile}. For NaNiO$_2$, preliminary analysis of NOMAD data indicated Na occupancy was 1 within error, hence site occupancy of all sites during all further refinement was fixed at 1. Thermal B$_\mathrm{eq}$ parameters were allowed to refine but constrained to be positive and not exceed a value of 5\,\AA$^2$. All atomic positions were refined within symmetry constraints. The background was fitted by a Chebyschev polynomial (order 6 for PEARL data, order 11 for NOMAD data, order 19 for XRD data). For XRD data, a TCHZ peak-shape was used~\cite{thompson1987rietveld}. Peak-shapes used for neutron data are discussed in SI Section II. For PEARL only the 90$^\circ$ detection bank was used, but for NOMAD a combined refinement was performed using banks~2-5 ($2\theta=31^\circ,67^\circ,122^\circ,154 ^\circ$ respectively).
	
	\section{\label{sec:level4}Results}
	
	\subsection{Ambient-pressure structural properties.}
	
	Powder X-ray diffraction of the as-synthesised NaNiO$_2$ indicated the formation of a phase-pure product. SEM on the material indicates the sample is polycrystalline with particulates between 0.2~$\micro$m and 5\,$\micro$m in diameter [SI Figure S13]. Rietveld refinement using the reported monoclinic $C2/m$ space group [SI Figure~S1; Table~S1] yielded lattice parameters consistent with prior reports~\cite{dick1997structure,sofin2005new}. 
	
\begin{figure}[]
		\includegraphics[scale=1]{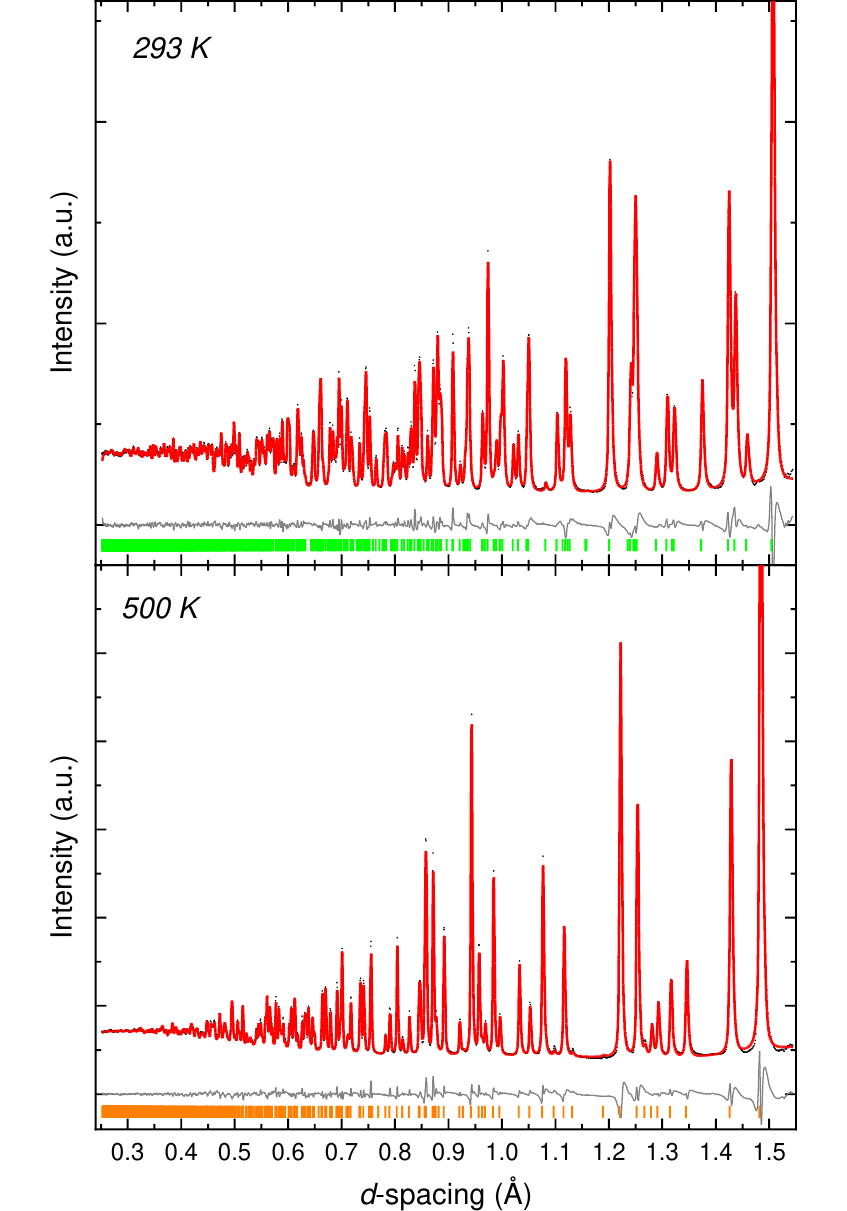}
		\caption{\label{SNS_data-example-Rietveld}Rietveld refinements for the ambient-pressure, variable-temperature neutron diffraction measurements of NaNiO$_2$ on bank 5 of NOMAD ($2\theta = 154^\circ$) at 293\,K (top) and 500\,K (bottom). Black dots: measured data; red line: calculated diffraction pattern from Rietveld refinement; grey line: $\mathrm{Y}_\mathrm{obs}-\mathrm{Y}_\mathrm{calc}$. Green and orange tick marks show expected reflections for the monoclinic and rhombohedral phase respectively.}
	\end{figure}

	The reported monoclinic$\Rightarrow$rhombohedral phase transition in NaNiO$_2$ was investigated using neutron powder diffraction at ambient pressure on the NOMAD instrument. Rietveld refinement [Figure~\ref{SNS_data-example-Rietveld}] shows the phase transition occurs between 450\,K and 500\,K, and is reversible on cooling. The lattice parameters [SI Table~S2] all exhibit positive thermal expansion, and are consistent with previous measurements~\cite{dick1997structure,chappel2000study}. 
	
	In the monoclinic structure the NiO$_6$ octahedra exhibit a cooperative JT distortion with 2 longer Ni-O bonds, whereas in the high-temperature rhombohedral phase all six Ni-O bonds are equivalent [Figure~\ref{struc_fig}]. The degree of bond length distortion within individual NaO$_6$ and NiO$_6$ octahedra can be evaluated using a number of distortion metrics, calculated using \textsc{topas 5}~\cite{coelho2018topas}. Here we consider the effective coordination number~\cite{hoppe1979effective} and the bond length distortion indices~\cite{baur1974geometry}, which measure distortion in octahedra by quantifying the difference from the average value of the distances between the central cation and the coordinated oxygen anions. The general form of effective coordination, $\mathrm{ECoN}$, and bond length distortion index, $D$, is given in the SI. The equations as applicable to monoclinic NaNiO$_2$ are:
		
		\begin{equation}
		\mathrm{ECoN} = 4\exp{
			\left[ 1-\left(\frac{l_\mathrm{short}}{l'_\mathrm{av}}\right)^6\right]}
		+ 2\exp{
			\left[ 1-\left(\frac{l_\mathrm{long}}{l'_\mathrm{av}}\right)^6\right]},
		\end{equation}
		
		where $l'_\mathrm{av}$ is a modified average bond length defined in SI, and;
		
		\begin{equation}
		D=\frac{1}{3}\frac{l_\mathrm{long} - l_\mathrm{av}}{l_\mathrm{av}} + \frac{2}{3}\frac{l_\mathrm{av} - l_\mathrm{short}}{l_\mathrm{av}}
		\end{equation}
		
		where $l_\mathrm{av}$ is the average bond length, and $l_\mathrm{long}$ and $l_\mathrm{short}$ are the long and short bonds respectively. 
		
	In the rhombohedral structure the effective coordination number and bond length distortion index of both the NiO$_6$ and NaO$_6$ octahedra are constrained by symmetry to values of 6 and 0 respectively. In the monoclinic structure, departure from these values indicates bond length disproportionation and is primarily attributable to the JT distortion. These changes are significantly larger for the JT-active NiO$_6$ octahedra than the NaO$_6$ octahedra. 
	%Throughout the measurement, the effective coordination of NaO$_6$ octahedra remains very near its high-symmetry value of 6, for example at 293\,K the value of effective coordination in NaO$_6$ octahedra is 5.99232(19), compared with 5.309(3) in the NiO$_6$ octahedra [SI Table~S7]. This is indicative of much greater distortion in bond lengths for NiO$_6$ octahedra, consistent with the JT distortion. Similarly, the bond length distortion index is an order of magnitude higher for NiO$_6$, at 0.05463(14), than NaO$_6$ octahedra at 0.00581(11) at 293\,K.
	%Throughout the measurement, the effective coordination/bond length distortion index of NaO$_6$ octahedra remains very near its high-symmetry value of 6/0, for example at 293\,K the value of effective coordination/bond length distortion index in NaO$_6$ octahedra is 5.99232(19)/0.00581(11), compared with 5.309(3)/0.05463(14) in the NiO$_6$ octahedra [SI Table~S7]. This is indicative of much greater distortion in bond lengths for NiO$_6$ octahedra, consistent with the JT distortion.
	%Throughout the measurement, the bond length distortion index (effective coordination) of NaO$_6$ octahedra remains very near its high-symmetry value of 0(6), for example at 293\,K the value of bond length distortion index (effective coordination) in NaO$_6$ octahedra is 0.00581(11)(5.99232(19)), compared with 0.05463(14)(5.309(3)) in the NiO$_6$ octahedra [SI Table~S7]. This is indicative of much greater distortion in bond lengths for NiO$_6$ octahedra, consistent with the JT distortion.
	Throughout the measurement, the bond length distortion index $\{\mathrm{effective~coordination}\}$ of NaO$_6$ octahedra remains very near its high-symmetry value of 0$\{6\}$, for example at 293\,K the value of bond length distortion index $\{\mathrm{effective~coordination}\}$ in NaO$_6$ octahedra is 0.00581(11)$\{5.99232(19)\}$, compared with 0.05463(14)$\{5.309(3)\}$ in the NiO$_6$ octahedra. This is indicative of much greater distortion in bond lengths for NiO$_6$ octahedra, consistent with the JT distortion. The values of bond length distortion index are on the same order of magnitude as recent studies on JT-distorted Mn$^{3+}$O$_6$-containing compounds~\cite{kimber2012charge,lawler2021decoupling}.
	
	Inconsistency in bond length is not the only distortion of the octahedra from regular octahedra. A regular octahedron would have bond angles $\theta_{\mathrm{O-}M\mathrm{-O}} = 90^\circ$ for nearest-neighbour O anions. However, in both the JT-active monoclinic phase and the JT-inactive rhombohedral phase there is variance from this ideal bond angle. Non-nearest-neighbour oxygen anions are constrained to have 180$^\circ$ bond angles via the central cation, and so the 12 bond angles in an octahedron are each paired with another O-$M$-O bond, with the paired bond angles sharing one oxygen in common and with their non-shared oxygen anions occurring along a straight line through the central cation (for a visual representation, see SI Figure~S14). We define these bond angles as $\theta_{\mathrm{O-}M\mathrm{-O}} = 90^\circ \pm \Delta$, where the two angles in a pair have opposite sign preceding the $\Delta$. $\Delta$ can also be thought of as a measure of the extent of angular distortion. In the rhombohedral structure, there is only one value of $\Delta$ for each type of octahedron, with half the O-$M$-O bond angles being $90^\circ + \Delta$ and the other half being $90^\circ - \Delta$. In the monoclinic unit cell where octahedra have two long $M$-O ($M$=Na,Ni) bonds and four short $M$-O bonds, there are four nearest-neighbour bond angles between short and short bonds and eight nearest-neighbour bond angles between short and long bonds. We therefore must define two values of $\Delta$ for the bond angles in the monoclinic phase, $\Delta_\mathrm{short-short}$ and $\Delta_\mathrm{long-short}$, respectively. Table~\ref{angle_table} shows these values of $\Delta$ at each temperature. It is clear that NaO$_6$ octahedra exhibit far higher bond angle distortion than NiO$_6$ octahedra, in contrast to the bond length distortion which is greater for NiO$_6$ octahedra. This is not unexpected, given that crystal field effects will result in much greater stability for open-shell $d^7$ Ni$^{3+}$ in an octahedral configuration, minimising bond angle variance, whereas this won't be a factor for closed-shell Na$^+$ cations.
	
	\begin{table}[]
		\begin{tabular}{c | c | c c | c c}
			\toprule
			\multirow{ 2}{*}{Phase} & \multirow{ 2}{*}{T (K)}   & \multicolumn{2}{c}{NiO$_6$ ($^\circ$)} & \multicolumn{2}{c}{NaO$_6$ ($^\circ$)}\\ 
			&    & $\Delta^\mathrm{Ni}_\mathrm{short-short}$  & $\Delta^\mathrm{Ni}_\mathrm{long-short}$ & $\Delta^\mathrm{Na}_\mathrm{short-short}$  & $\Delta^\mathrm{Na}_\mathrm{long-short}$ \\ 
			\midrule
			$C2/m$ & 293 (-) & 6.134(17)   & 5.456(19)  & 14.494(12) & 9.708(14) \\
			$C2/m$ & 450 ($\uparrow$) & 6.163(19) & 5.50(2) & 14.564(14) & 9.844(16) \\
			$R\bar{3}m$ & 500 ($\uparrow$) & \multicolumn{ 2}{c}{6.135(15)}  & \multicolumn{ 2}{c}{11.777(13)}   \\
			$C2/m$ & 316 ($\downarrow$) & 6.121(18) & 5.46(2) & 14.505(13) & 9.731(15) \\
			\bottomrule
		\end{tabular}
		\caption{\label{angle_table}This table shows the value of $\Delta$ for bond angles O-$M$-O ($M$=Na, Ni) which take the value $90^\circ\pm\Delta$, as a function of temperature. For definitions, check the main text. The arrow next to the temperature indicates whether the data were collected on warming or cooling of the sample.}
	\end{table}
		
	\subsection{Variable-pressure neutron diffraction.}
	
\begin{figure}[]
		\includegraphics{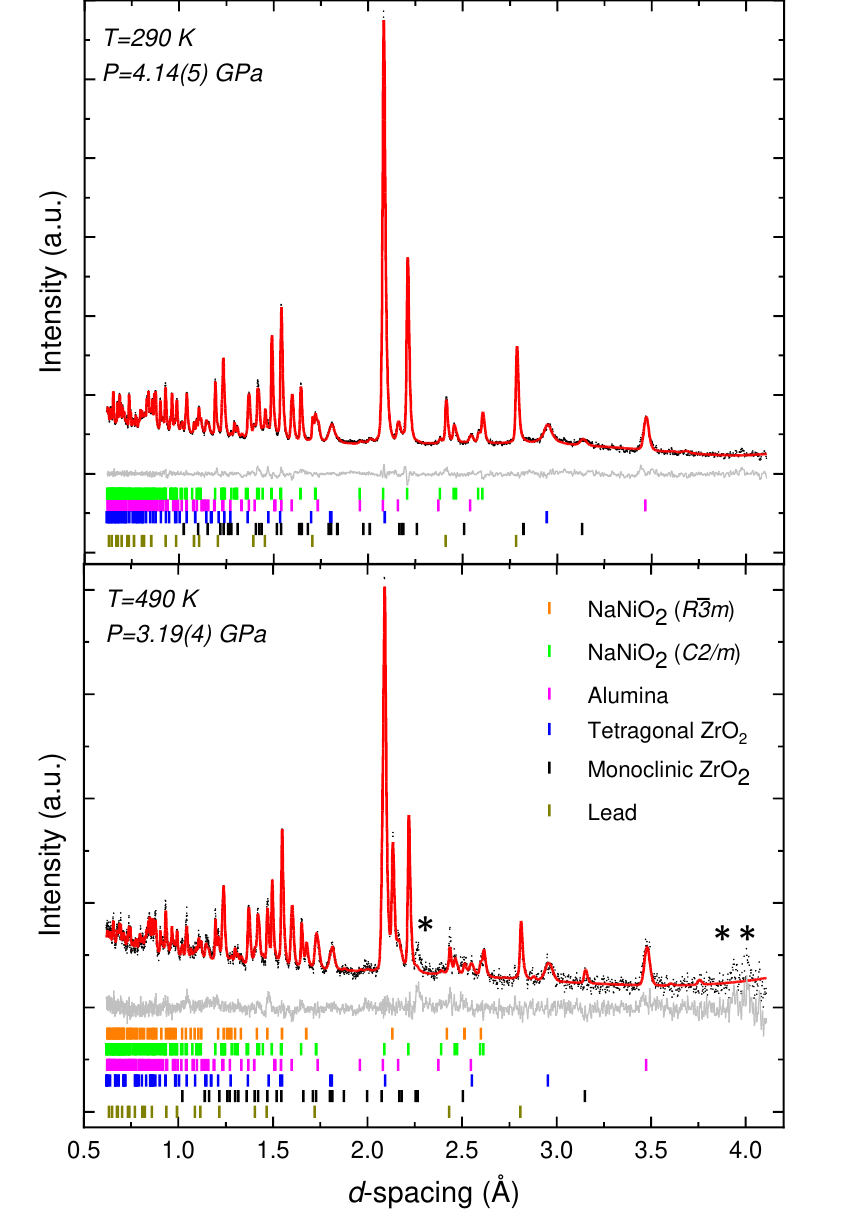}
		\caption{\label{Rietveld_fits_PEARL}Rietveld refinements for the variable-pressure neutron diffraction data of NaNiO$_2$. Top: a representative plot with monoclinic NaNiO$_2$ only; bottom: a representative plot with both monoclinic and rhombohedral NaNiO$_2$. Black dots: measured data; red line: calculated diffraction pattern from Rietveld refinement; grey line: $\mathrm{Y}_\mathrm{obs}-\mathrm{Y}_\mathrm{calc}$. %Tickmarks are displayed at the bottom of the peaks to show which peaks are due to which phase. 
			Unfitted peaks are marked with an asterisk and arise from crystalline fluorinert [SI Figure~S3].}
	\end{figure}

\begin{figure}[hbtp]
		\includegraphics{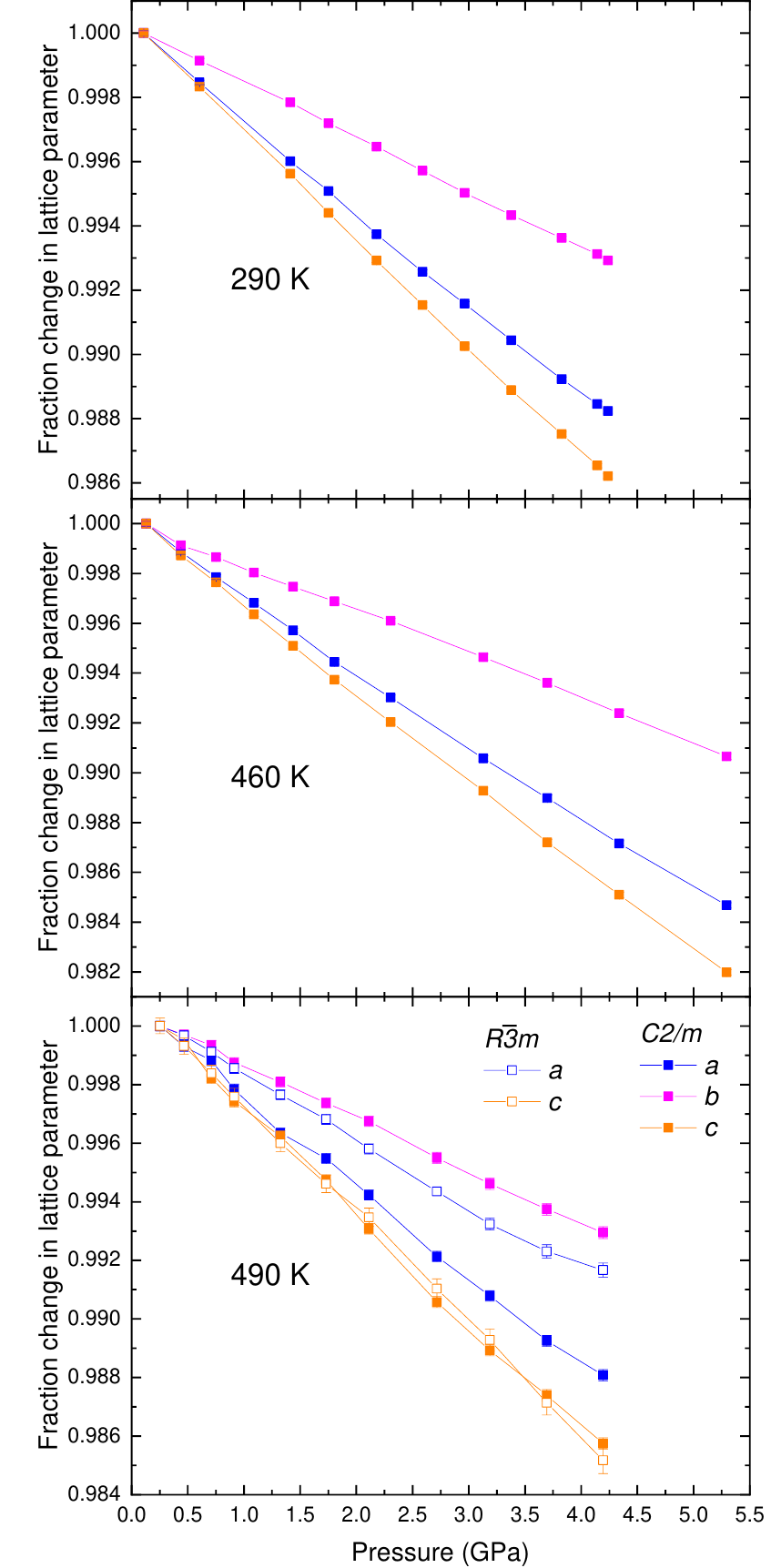}
		\caption{\label{lattice_params_percent}Fractional contraction of lattice parameters obtained by Rietveld refinement~\cite{rietveld1969profile} as a function of temperature and pressure for the monoclinic $C2/m$ and rhombohedral $ R\bar{3}m$ phases of NaNiO$_2$. Error bars are shown for all plots; where they are not visible it is because the error is smaller than the data point. Lines are a guide for the eye.}
	\end{figure}

\begin{figure}[t]
		\includegraphics{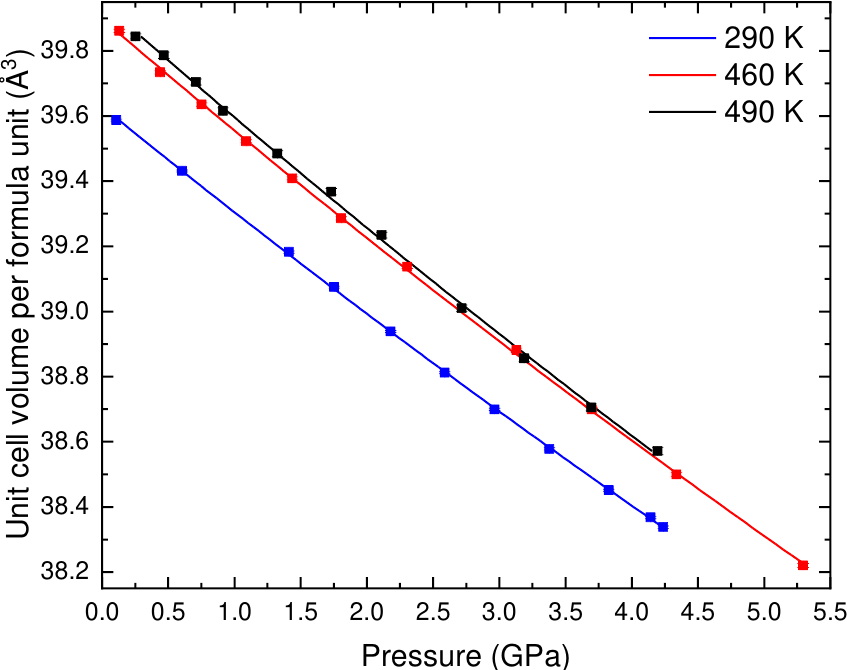}
		\caption{\label{lattice_params_fig}Variation in unit cell volume per formula unit for monoclinic $C2/m$ phase. Solid data points show experimental derived values and the solid line the determined 2nd-order Birch-Murnaghan equation of state. Full lattice parameters are found in SI Tables~S3-5. Where error bars are not visible it is because they are smaller than the data points.}
	\end{figure}

\begin{figure}[hbtp]
		\includegraphics{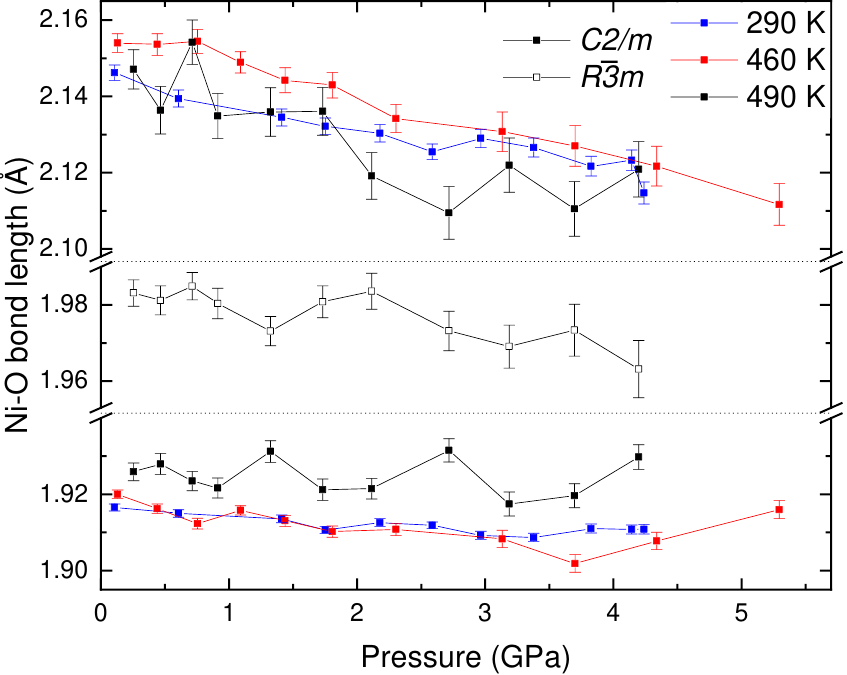}
		\caption{\label{bond_lengths}Ni-O bond lengths, as a function of pressure, and associated error of monoclinic NaNiO$_2$ at 290\,K, 460\,K, and 490\,K, with the JT-inactive rhombohedral phase bond lengths shown for 490\,K. %Where error bars are not shown, it is because they are smaller than the data point.
			Lines are a guide for the eye.}
	\end{figure}

\begin{figure}[hbtp]
		\includegraphics{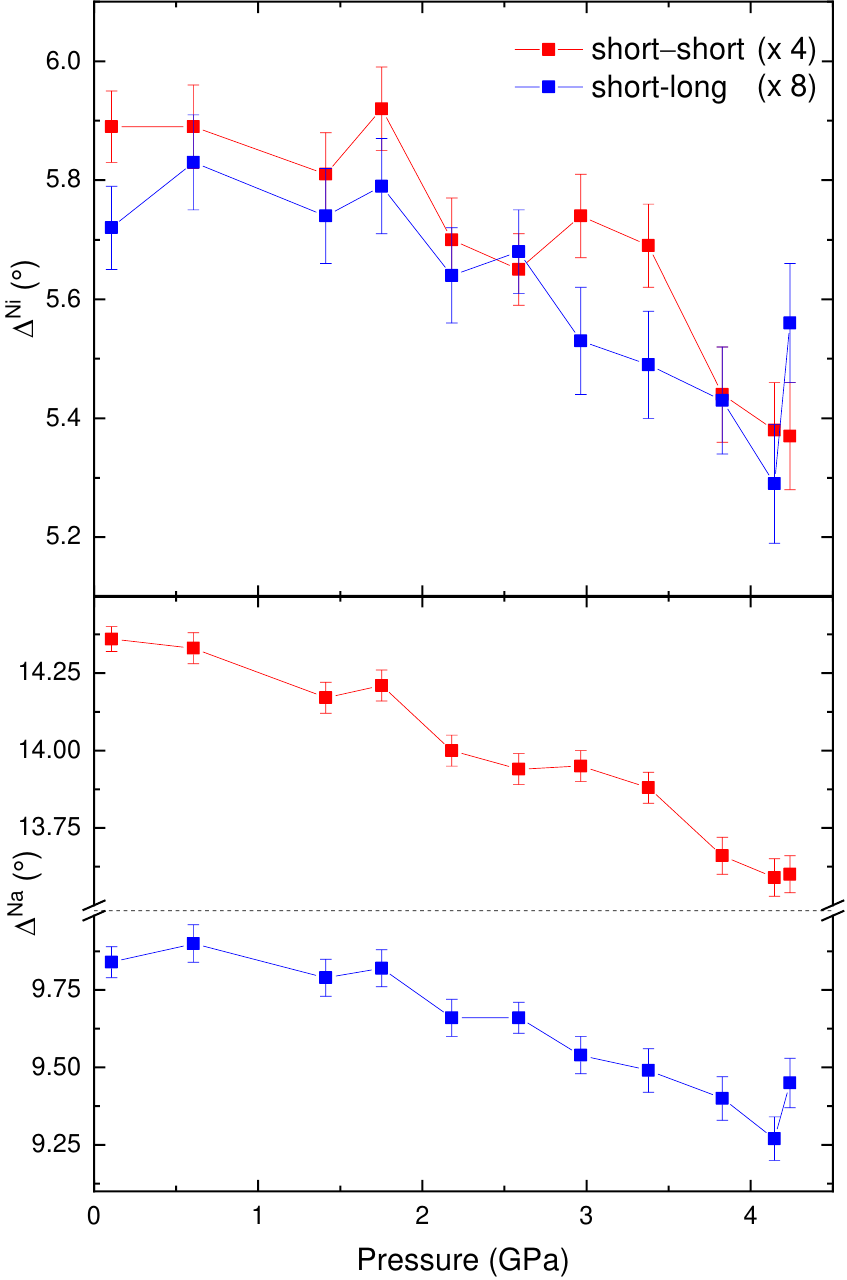}
		\caption{\label{bond_angles}Values of $\Delta$ for NiO$_6$ and NaO$_6$ octahedra as a function of pressure at 290\,K, representing the magnitude of angular distortion as nearest-neighbour bond angles take the value $90^\circ \pm \Delta$. The two different $\Delta$ values in monoclinic NaNiO$_2$ are between two short bonds (red) and between a short and long bond (blue), where bonds are short or long due to the JT distortion. Lines are a guide for the eye.}
	\end{figure}

\begin{figure}[bt]
		\includegraphics{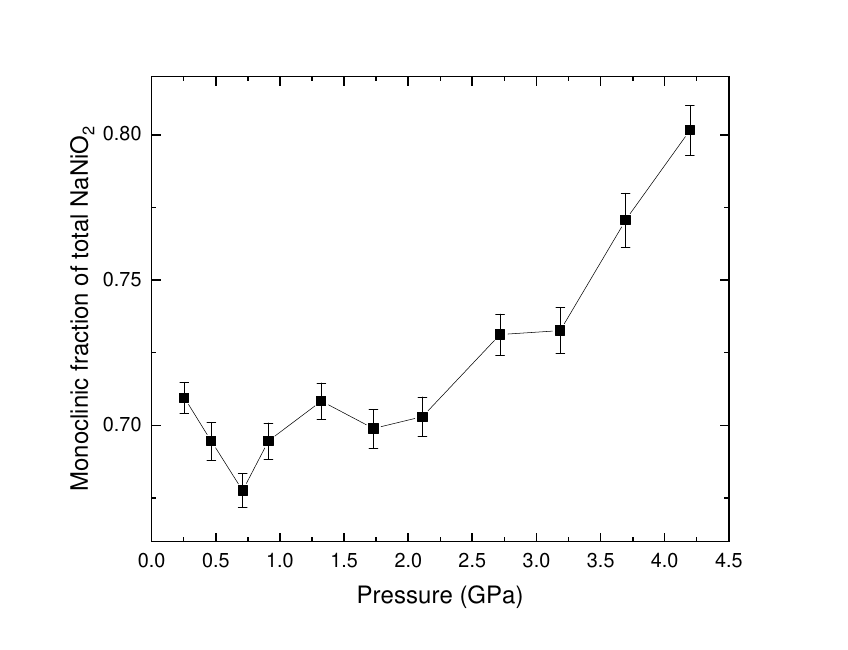}
		\caption{\label{mono_frac}The fraction of NaNiO$_2$ which is in the monoclinic phase at 490\,K, as a function of pressure.}
	\end{figure}

	The effect of pressure on the JT distortion in NaNiO$_2$ was explored at 290\,K, 460\,K, and 490\,K, with an example Rietveld refinement shown in Figure~\ref{Rietveld_fits_PEARL} and data shown in Figures~\ref{lattice_params_percent}-\ref{mono_frac}. Over the entire pressure and temperature range studied NaNiO$_2$ could be described using the previously-reported ambient pressure crystal structures. Diffraction data also included contributions from alumina and zirconia in the sample environment and the lead used to determined the applied pressure - these are also included in the structural refinements. In addition, at higher temperatures (460\,K and 490\,K) and pressures additional peaks attributed to crystallisation of the fluorinert pressure media [SI Figure~S3] are observed in the measurements. 
	
	Rietveld analysis, Figure~\ref{Rietveld_fits_PEARL}, shows that NaNiO$_2$ remained in the monoclinic phase at 290\,K (up to 4.24(5)\,GPa) and 460\,K (up to 5.29(8)\,GPa). However, the measurements at 490\,K capture NaNiO$_2$ midway through its transition from JT-distorted $C2/m$ monoclinic to JT-inactive $ R\bar{3}m$ rhombohedral, and throughout this isotherm the NaNiO$_2$ is mixed-phase.
			
	The lattice parameters [Figure~\ref{lattice_params_percent}] show the expected variation with temperature and pressure. The rhombohedral and monoclinic phases have similar compressibility, and in both NaNiO$_2$ is considerably more compressible in the inter-layer direction ($c$-axis) than the intra-layer ($ab$-) plane. With reference to Figure~\ref{struc_fig}, we note that compression within the plane results in changes to the highly ionic Na$^+$-Na$^+$ interactions and the less ionic, but still repulsive Ni$^{3+}$-Ni$^{3+}$ interactions, whereas compression in the inter-layer direction will compress the Ni$^{3+}$-O$^{2-}$ and Na$^{+}$-O$^{2-}$ bonds which are softer due to the nearest-neighbour interaction lacking a Coulomb repulsive force. This higher compressibility in the inter-layer direction is consistent with that seen in another material with alternating layers of edge-sharing octahedra, the honeycomb iridate Na$_2$IrO$_3$~\cite{hermann2017high,layek2020electronic}.
	
	Within the plane, in monoclinic NaNiO$_2$, the $b$-axis is less compressible than the $a$-axis. 
	A reason for this might be that Na$^+$-Na$^+$ and Ni$^{3+}$-Ni$^{3+}$ interactions are parallel to the direction of compression for the $b$-axis, maximising the increase in Coulomb repulsion with decreasing lattice parameter due to compression, whereas there are no Na$^+$-Na$^+$/Ni$^{3+}$-Ni$^{3+}$ interactions with components only along the $a$-axis. Another contribution may be that the Na$^+$-Na$^+$ and Ni$^{3+}$-Ni$^{3+}$ ionic distances parallel to the $b$ axis are considerably shorter than the distances which can be projected onto the $a$-axis 
	($\sim$2.85%2.84614(11)
	\,\AA~and $\sim$3.02%3.01661(12)
	\,\AA~respectively at 290\,K and 0.107(8)\,GPa).
	
	%For the high-temperature, rhombohedral phase, Figure~\ref{lattice_params_percent}(c) shows that the $c$ axis has approximately the same compressibility as the $c$ axis for the monoclinic phase. The rhombohedral $a$ parameter is the unit cell length in both the $a$ and $b$ directions, as the removal of the monoclinic distortion increases the symmetry of the unit cell. The compressibility of the rhombohedral $a$ parameter is less than the monoclinic $a$ but more than the monoclinic $b$. 
	
	\begin{table}[]
		\begin{tabular}{c c c c}
			\toprule
			Structure & Temperature & $V_0$ ($\AA^3$) & $B_0$ (GPa)   \\ \midrule
			$ R\bar{3}m$  & 490\,K  & 119.83(2) & 113(1)    \\ \midrule
			$C2/m$    & 490\,K  & 79.900(16) & 110(1) 	 \\ 
			& 460\,K  & 79.798(9) & 113.5(6) 	 \\ 
			& 290\,K &   79.258(7) & 119.6(5)    \\ 
			\bottomrule
		\end{tabular}
		\caption{\label{table_BM}Parameters determined from the 2nd-order Birch-Murnaghan Equation of State, obtained using PASCal~\cite{cliffe2012pascal} [SI Section~V, Table~S12].}
	\end{table}
	
	\mbox{PASCal}~\cite{cliffe2012pascal} was used to obtain the bulk modulus for each isotherm, using a 2nd-order Birch-Murnaghan equation of state~\cite{birch1947finite}. A plot of the unit cell volume obtained by Rietveld refinement, as a function of pressure, with a fit of this equation of state, is shown in Figure~\ref{lattice_params_fig} and tabulated in Table~\ref{table_BM}. For the monoclinic phase, $\frac{dV_0}{dT} > 0$ which is consistent with a structure with positive thermal expansion. $B$ decreases with increasing temperature, meaning that compressibility increases with temperature. At 290\,K, $B$ is 119.6(5)\,GPa. This is comparable with a similar JT-distorted material with edge-sharing octahedra, CuMnO$_2$, which has bulk modulus 116(2)\,GPa~\cite{lawler2021decoupling}. It is, however, substantially less than the reported bulk modulus for ZnMn$_2$O$_4$ of 197(5)\,GPa~\cite{aasbrink1999high}, and although there are several different reported values for Mn$_3$O$_4$ depending on phase and morphology~\cite{ovsyannikov2021structural,paris1992mn3o4,li2020size}, all are higher than what we report for NaNiO$_2$. LaMnO$_3$ is not entirely comparable owing to the LaO$_{12}$ units and corner-sharing octahedra, but for reference it has a reported bulk modulus of 108(2)\,GPa~\cite{loa2001pressure}. 
	
	The directions of the principal axes of compression are determined using PASCal~\cite{cliffe2012pascal} [SI Figure~S13]. These are the axes in which compression occurs linearly with pressure, and do not necessarily align with the crystallographic axes in crystalline materials. The principal axis directions projected onto the $ab$-plane do not change between the monoclinic and rhombohedral phases. However, the inter-layer direction is a principal axis for the rhombohedral phase, but not for the monoclinic phase where two principal axes are at an angle to the inter-layer direction [Fig.~\ref{struc_fig}]. Interestingly, the axis of JT elongation does not correspond to any of the principal axes [SI Table~S11]. There is some temperature dependence in the principal axis directions, likely owing to variation in the lattice parameters with temperature [SI Tables.~S3-S5]. The compressibilities of NaNiO$_2$ in each of the principal axes is consistent with variation in lattice parameters [SI Table~S11]. 
		
	We now consider pressure-dependence of the bond length distortion index and effective coordination [SI Figure~S8; Tables~S8-10]. As in the ambient pressure measurements the bond length distortions are significantly larger in the NiO$_6$ compared to the NaO$_6$ octahedra, with the most significant variation being the increase in effective coordination (5.387(10) at 0.107(8)\,GPa to 5.504(13) at 4.24(5)\,GPa at 290\,K) and decrease in bond length distortion index (from 0.0512(5) to 0.0458(7) in the same pressure range at 290\,K) of the NiO$_6$ octahedra in the monoclinic phase on application of pressure. NaO$_6$ octahedra exhibit far smaller changes in bond length distortion index and effective coordination, with the overall behaviour not seeming to exhibit a consistent change with pressure; effective coordination remains between 5.98 and 5.99 throughout the 290\,K isotherm. 
	
	The differing behaviour of bond length distortion index between NiO$_6$ and NaO$_6$ octahedra is likely attributable to the fact that NiO$_6$ is JT-active and NaO$_6$ is not, and suggests that pressure is decreasing the magnitude of JT distortion. We investigate this by considering the direct manifestation of the JT effect in NaNiO$_2$. The Ni-O bond lengths of both the monoclinic and rhombohedral phases as a function of pressure are shown in Figure~\ref{bond_lengths}. The short Ni-O bonds are less sensitive to the effect of pressure than the long Ni-O bonds, indicating that the difference between long and short Ni-O bond lengths is decreasing with pressure. We also observe that the average monoclinic bond length is consistently larger than the rhombohedral bond length at 490 K [SI Figure~S9]. In the NaO$_6$ octahedra, SI Figure~S5, there is an approximately linear variation of the Na-O bond lengths with pressure. We conclude that anisotropy of the Ni-O bond compression is a consequence of the JT distortion in NiO$_6$ octahedra. 
	
	The observed decrease in difference between long and short Ni-O bonds with pressure is also reported for other materials containing a JT distortion, such as LaMnO$_3$~\cite{loa2001pressure}, KCuF$_3$~\cite{zhou2011jahn}, and CuAs$_2$O$_4$~\cite{caslin2016competing}. This is equivalent to the observed tendency with pressure of NiO$_6$ octahedra bond length distortion index and effective coordination towards their symmetry-constrained values of 0 and 6, respectively. It indicates that the symmetry of JT-distorted octahedra increases with application of pressure in monoclinic NaNiO$_2$, consistent with prior reports~\cite{loa2001pressure,zhou2011jahn,caslin2016competing,collings2018disorder,bostrom2019high}.
	
	A previous study on LaMnO$_3$ attempted to extrapolate a linear fit to the pressure-dependence of JT-distorted bond length, and estimated a critical pressure of $\sim$18\,GPa~\cite{loa2001pressure}. Such an extrapolation could be performed for NaNiO$_2$ yielding a critical pressure of $\sim$50\,GPa, converging at a Ni-O bond length of 1.85\,\AA~at 290\,K. However, this value is unlikely to be representative of the true critical pressure of the JT distortion in NaNiO$_2$. A later study on LaMnO$_3$ found that the JT distortion was suppressed at a lower pressure of around $\sim$12\,GPa, suggesting such extrapolation does not yield accurate predictions. Further, studies of other JT-distorted materials such as [(CH$_3$)$_2$NH$_2$][Cu(HCOO)$_3$]~\cite{collings2018disorder} and CuMnO$_2$~\cite{lawler2021decoupling} have found that this pressure-dependence of JT-disproportionated bond length exists only up to a certain pressure, beyond which there is a change in behaviour which renders such extrapolation of low-pressure behaviour meaningless.
	
	We earlier defined the bond angles $\theta^\mathrm{M}_\mathrm{short-short}$ and $\theta^\mathrm{M}_\mathrm{long-short}$ ($M$=Na,Ni) for monoclinic NaNiO$_2$, and the associated $\Delta$ values which reduce the number of parameters needed to describe the behaviour. We plot these $\Delta$ values in Figure~\ref{bond_angles} for the 290\,K isotherm. These plots show that throughout the studied pressure range, the degree of angular distortion is far greater for NaO$_6$ than NiO$_6$, as was the case at ambient pressure [Table~\ref{angle_table}]. We can also see that with application of pressure, $\Delta$ is decreasing in value; this indicates increasing symmetry towards the 90$^\circ$ degree bond angle for a perfect octahedron, analogous to the increasing symmetry with pressure we see with bond length distortion index. 
	
	The pressure dependence of the NaO$_6$ and NiO$_6$ octahedral volume in NaNiO$_2$ [SI Figure~S7] show that the changes in volume displays different pressure-dependence for NiO$_6$ octahedra and NaO$_6$ octahedra, as compared with the unit cell. NaO$_6$ octahedra have higher relative compressibility than the entire unit cell, and NiO$_6$ octahedra are much more resistant to compression. It has been shown that for perovskites with $A$O$_{12}$ and $B$O$_6$ polyhedra the parameters $M_A$ and $M_B$ can be used to predict the relative compressibility of the polyhedra via $\beta_B / \beta_A = M_A / M_B$, in which $\beta_i = - \frac{1}{R_{i}} \frac{dR_i}{dP}$ is the bond compressibility, $R_i$ is the distance between the central cation and the \textit{i}th O anion, and $M_i$ is a bond-valence parameter defined in SI~\cite{zhao2004new}. We apply this model to NaNiO$_2$ and find that $M_\mathrm{Ni} > M_\mathrm{Na}$ throughout the 290\,K isotherm [SI Figure~S12]. Accounting for the different values of $R_i$, this indicates that $\frac{dR_\mathrm{Na-O}}{dP} > \frac{dR_\mathrm{Ni-O}}{dP}$, which is consistent with our observation that NaO$_6$ octahedra are more compressible than NiO$_6$ octahedra. This may be due to differences in electronic configuration for closed-shell Na$^+$ and open-shell Ni$^{3+}$, or Na$^+$ being a much larger ion than Ni$^{3+}$.
	
	We now consider a related model proposed by Angel \textit{et al.}, again in the context of perovskites~\cite{angel2005general}, whereby a transition temperature $T_\mathrm{c}$ associated with an octahedral phase transition will exhibit $dT_\mathrm{c}/dP < 0$ if octahedra are more compressible than the extra-framework cation sites (analogous to the NaO$_6$ octahedra in NaNiO$_2$), and $dT_\mathrm{c}/dP > 0$ if octahedra are less compressible. Our structural analysis shows the enhanced compressibility of NaO$_6$ octahedra when compared to NiO$_6$, and so this model predicts the observed increase in $T_\mathrm{JT}$ with pressure. It is worth noting that there are more degrees of freedom in the layered NaNiO$_2$ structure so the relationships between distortions in NiO$_6$ and NaO$_6$ may not be so strongly coupled as in the perovskites. However, the basic hypothesis of the model in Angel \textit{et al.} appears to be applicable to NaNiO$_2$.
	
	Along the 490\,K isotherm, both monoclinic and rhombohedral NaNiO$_2$ were observed in coexistence. The fraction of NaNiO$_2$ in the low-temperature, JT-distorted monoclinic phase is shown in Figure~\ref{mono_frac}. The fraction remains approximately stagnant to $\sim$2\,GPa, beyond which it consistently increases with increasing pressure. In the range where it is increasing, the monoclinic fraction at 490\,K increases from 67.8(6)\% at 0.71(2)\,GPa to 80.2(9)\% at 4.20(6)\,GPa. This indicates that $T_\mathrm{JT}$ increases with increasing pressure beyond $\sim$2\,GPa, consistent with our prediction based on octahedral compressibility.
	
	To explore the P-T dependence of the transition, the sample was heated at 5.29(8)\,GPa from 460\,K to 490\,K after measuring the variable-pressure 460\,K isotherm. At ambient pressure, this would result in a mixed monoclinic/rhombohedral phase. However, at the resulting high pressure of 5.46(9)\,GPa, we did not observe emergence of any rhombohedral peaks in the diffraction pattern. Subsequent reduction in pressure to 0.342(13)\,GPa at the same temperature, 490\,K, did yield emergence of rhombohedral peaks [Figure~S4], further supporting our interpretation that $T_\mathrm{JT}$ is increasing with pressure.%This suggests $T_\mathrm{JT}$ is higher at high pressure than at low pressure.  
	
\section{Discussion}
	
\begin{figure}[tb]
		\includegraphics[trim=0.6cm 0.5cm 1cm 1.5cm,clip,width=86mm]{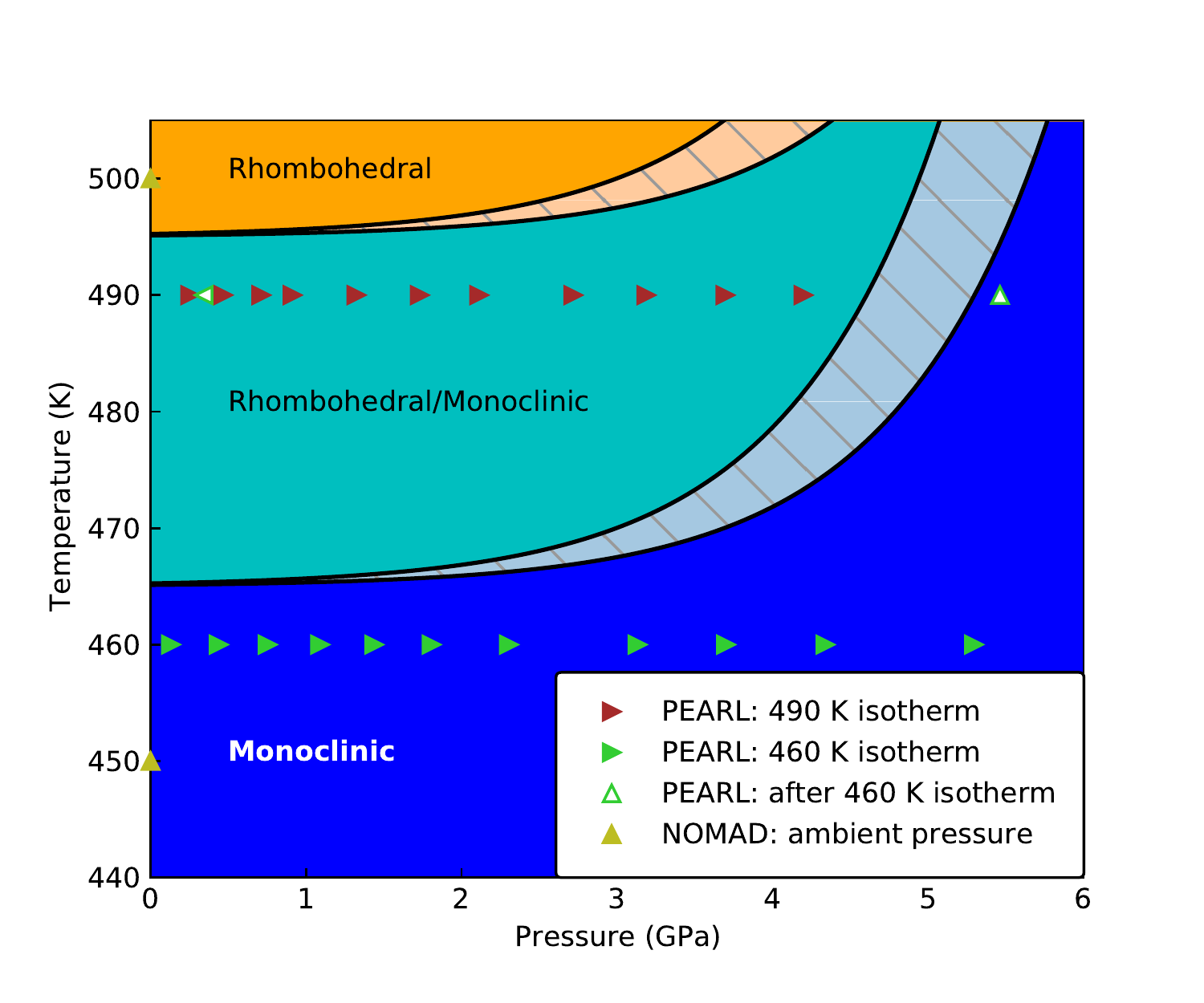}
		\caption{\label{PhaseDiagram.pdf}Tentative phase diagram showing the structure of NaNiO$_2$ as a function of pressure and temperature. Triangles denote diffraction measurements and point left/right if $P$ was decreasing/increasing or up/down if $T$ was increasing or decreasing.
			%presents the temperature and pressure points measured in this study and whether this corresponds to a rhombohedral, monoclinic, or mixed phase. 
			The precise boundaries of the three regions are estimates based on available data, with the results in Refs.~\cite{sofin2005new,chappel2000study} used to estimate the broadness of the transition.
		}
	\end{figure}

	The results of our $P$-\textit{T} study on NaNiO$_2$ are summarised in a phase diagram, Figure~\ref{PhaseDiagram.pdf}. To our knowledge, this is the first study on the effect of pressure on the JT transition temperature in a material containing edge-sharing \textit{M}O$_6$ octahedra, and the first variable-pressure study on a JT-active edge-sharing nickelate. Comparison between the results of this study and previous works must therefore rely on the work done on non-nickelate materials.
	
	Like the perovskite materials LaMnO$_3$~\cite{loa2001pressure,pinsard2001stability} and KCuF$_3$~\cite{zhou2011jahn}, NaNiO$_2$ exhibits far greater compressibility in the JT-elongated O-Ni-O axis than the JT-compressed O-Ni-O axes, with the JT distortion in both NaNiO$_2$ and the previously-discussed perovskites decreasing in magnitude with pressure. The consistent behaviour with other JT-active materials is also clear evidence that the charge disproportionation model proposed for LiNiO$_2$~\cite{chen2011charge,foyevtsova2019linio,green2020evidence} and some Ni$^{3+}$-containing perovskites~\cite{johnston2014charge} is not applicable to NaNiO$_2$.
	
	A novel behaviour we observe in NaNiO$_2$ is that $T_\mathrm{JT}$ increases with application of pressure NaNiO$_2$. For comparison, in LaMnO$_3$ the JT distortion is suppressed at $\sim$12\,GPa, indicating that $T_\mathrm{JT}$ is reduced to room temperature from $\sim$750\,K by 12\,GPa~\cite{zhou2008breakdown}. This mechanism seems unlikely in NaNiO$_2$ due to the increasing $T_\mathrm{JT}$ with pressure, although we cannot exclude the possibility that a reversal above our maximum measured pressure may result in a decrease in $T_\mathrm{JT}$. Additionally, there is a trend observed in this, and other~\cite{loa2001pressure,zhou2011jahn,caslin2016competing,collings2018disorder,bostrom2019high}, works, that the magnitude of distortion due to the JT effect decreases with pressure. This could be interpreted as meaning that there is some pressure where the distortion is entirely suppressed and the NiO$_6$ octahedra achieve a bond length distortion index of zero, consistent with absence of an ordered JT distortion. However, this is at odds with recent reports~\cite{collings2018disorder,lawler2021decoupling} which show that at some pressure the long and short bonds in JT-distorted octahedra eventually stabilise at different lengths. It is therefore not clear how exactly the JT distortion is suppressed in NaNiO$_2$ with high pressure, and further investigation is needed to elucidate this.
		
We should once again note that the conductivity behaviour of the high-temperature phase of NaNiO$_2$ also remains unexplored. There is significant reduction in resistivity with JT suppression in LaMnO$_3$,~\cite{zhou1999paramagnetic} and there is a possibility for similar behaviour in high-temperature rhombohedral NaNiO$_2$. Density Functional Theory calculations on rhombohedral, JT-free LiNiO$_2$ (which is isostructural with the high-temperature phase of NaNiO$_2$) have suggested metallic behaviour~\cite{chen2011first}. On a similar note, a broad first-order transition between two structures with a group-subgroup relationship in SrCrO$_3$~\cite{ortega2007microstrain} featured coexistence of electronic phases. If the high-temperature phase of NaNiO$_2$ were indeed metallic, this metallic behaviour could explain why $dT_\mathrm{JT}/dP > 0$ in this material, as application of pressure may result in narrowing of Ni(3\textit{d})-O(2\textit{p}) bands, pushing the metal-to-insulator phase transition to higher and higher temperatures, and electronic phase coexistence could provide an explanation for the very broad nature of the transition.
	
	\section{\label{sec:level6}Conclusion}
		
	The key finding of this study is that in NaNiO$_2$, $T_\mathrm{JT}$ increases slightly with application of pressure while JT-distorted NiO$_6$ octahedra become more symmetric, as demonstrated by the pressure-dependence of two distortion metrics (effective coordination and bond length distortion index). While the latter is a well-documented property of JT-distorted materials, the former is in contrast to the JT distortion in LaMnO$_3$~\cite{zhou2008breakdown}. NaNiO$_2$ is more resistant to pressure than other similar materials, having a higher bulk modulus ($B = 119.6(5)$\,GPa at 290\,K) than similar perovskites~\cite{loa2001pressure,zhou2011jahn}, Prussian Blue analogues~\cite{bostrom2019high}, and layered honeycomb structures~\cite{hermann2017high}, although its bulk modulus is very similar to JT-distorted edge-sharing CuMnO$_2$~\cite{lawler2021decoupling}, and is less than Mn$_3$O$_4$~\cite{ovsyannikov2021structural,paris1992mn3o4,li2020size}, NiO~\cite{eto2000crystal}, and ZnMn$_2$O$_4$~\cite{aasbrink1999high}. NaNiO$_2$ also displays a much smaller magnitude of $\frac{dT_\mathrm{JT}}{dP}$ than LaMnO$_3$, with LaMnO$_3$ shifting $T_\mathrm{JT}$ from $\sim$750\,K at ambient pressure to room temperature in 12\,GPa~\cite{zhou2008breakdown} compared with a very small shift from $\sim$480\,K at ambient pressure in NaNiO$_2$. 
	
	Further variable-pressure diffraction measurements, at several temperatures and to higher pressures, are needed to fully understand the process of suppressing the JT distortion in NaNiO$_2$. Variable-pressure Raman spectroscopy measurements on NaNiO$_2$ could also be useful and may help identify phase transitions at  higher pressures.
	
	Additionally, future investigations are needed to investigate whether other JT-distorted materials exhibit a $dT_\mathrm{JT}/dP > 0$ pressure dependence, for example a study building on previous work on CuMnO$_2$~\cite{lawler2021decoupling} by measuring at multiple isotherms.

	\textbf{Data availability:} The variable-pressure neutron diffraction data from the PEARL instrument at ISIS is available at doi:10.5286/ISIS.E.RB2000219~\cite{https://doi.org/10.5286/isis.e.rb2000219}. All other data can be found at doi:10.17863/CAM.81605~\cite{https://doi.org/10.17863/CAM.81605}.
	
	\section*{Acknowledgements}
	This work was supported by the Faraday Institution grant number FIRG017. 
	LNC acknowledges a scholarship EP/R513180/1 to pursue doctoral research from the UK Engineering and Physical Sciences Research Council (EPSRC). 
	Experiments at the ISIS Neutron and Muon Source were supported by a beamtime allocation RB2000219 from the Science and Technology Facilities Council of the United Kingdom. 
	A portion of this research used resources at the Spallation Neutron Source, a DOE Office of Science User Facility operated by the Oak Ridge National Laboratory of the United States of America, with data collection performed by Joerg C Neuefeind and Jue Liu (ORNL).
	Figure~\ref{struc_fig} was generated using CrystalMaker\textsuperscript{\tiny\textregistered}: a crystal and molecular structures program from CrystalMaker Software Ltd, Oxford, UK.
	Heather Greer assisted with the SEM images in Supplementary Information.
	We also thank others whose ideas and comments were useful: Joshua D Bocarsly, Farheen N Sayed, Andrew G Seel, Siddharth S Saxena, Euan N Bassey, Nicola D Kelly, Venkateswarlu Daramalla, Chloe S Coates, Camilla Tacconis, and Debasis Nayak.
	
	%Magnetic measurements were carried out using the Advanced Materials Characterisation Suite at the Maxwell Centre in Cambridge, funded by EPSRC Strategic Equipment Grant EP/M000524/1.
	
	\section*{Supplementary Information}
	
	The Supporting Information is available free of charge at -------------. Additional neutron diffraction data and refinement details. SEM images. Tabulated diffraction data and distortion parameters. Transformation matrices.

	\bibliography{references}% Produces the bibliography via BibTeX.

\providecommand{\latin}[1]{#1}
\makeatletter
\providecommand{\doi}
  {\begingroup\let\do\@makeother\dospecials
  \catcode`\{=1 \catcode`\}=2 \doi@aux}
\providecommand{\doi@aux}[1]{\endgroup\texttt{#1}}
\makeatother
\providecommand*\mcitethebibliography{\thebibliography}
\csname @ifundefined\endcsname{endmcitethebibliography}
  {\let\endmcitethebibliography\endthebibliography}{}
\begin{mcitethebibliography}{61}
\providecommand*\natexlab[1]{#1}
\providecommand*\mciteSetBstSublistMode[1]{}
\providecommand*\mciteSetBstMaxWidthForm[2]{}
\providecommand*\mciteBstWouldAddEndPuncttrue
  {\def\EndOfBibitem{\unskip.}}
\providecommand*\mciteBstWouldAddEndPunctfalse
  {\let\EndOfBibitem\relax}
\providecommand*\mciteSetBstMidEndSepPunct[3]{}
\providecommand*\mciteSetBstSublistLabelBeginEnd[3]{}
\providecommand*\EndOfBibitem{}
\mciteSetBstSublistMode{f}
\mciteSetBstMaxWidthForm{subitem}{(\alph{mcitesubitemcount})}
\mciteSetBstSublistLabelBeginEnd
  {\mcitemaxwidthsubitemform\space}
  {\relax}
  {\relax}

\bibitem[Bhadram \latin{et~al.}(2021)Bhadram, Joseph, Delmonte, Gilioli,
  Baptiste, Godec, Lobo, and Gauzzi]{bhadram2021reentrant}
Bhadram,~V.; Joseph,~B.; Delmonte,~D.; Gilioli,~E.; Baptiste,~B.; Godec,~Y.~L.;
  Lobo,~R.; Gauzzi,~A. Reentrant phase transition and suppression of
  {J}ahn-{T}eller distortion in the quadruple perovskite structure under high
  pressure. \emph{arXiv preprint arXiv:2102.09998} \textbf{2021}, \relax
\mciteBstWouldAddEndPunctfalse
\mciteSetBstMidEndSepPunct{\mcitedefaultmidpunct}
{}{\mcitedefaultseppunct}\relax
\EndOfBibitem
\bibitem[Zhou \latin{et~al.}(2008)Zhou, Uwatoko, Matsubayashi, and
  Goodenough]{zhou2008breakdown}
Zhou,~J.-S.; Uwatoko,~Y.; Matsubayashi,~K.; Goodenough,~J. Breakdown of
  magnetic order in {M}ott insulators with frustrated superexchange
  interaction. \emph{Phys. Rev. B} \textbf{2008}, \emph{78}, 220402\relax
\mciteBstWouldAddEndPuncttrue
\mciteSetBstMidEndSepPunct{\mcitedefaultmidpunct}
{\mcitedefaultendpunct}{\mcitedefaultseppunct}\relax
\EndOfBibitem
\bibitem[Loa \latin{et~al.}(2001)Loa, Adler, Grzechnik, Syassen, Schwarz,
  Hanfland, Rozenberg, Gorodetsky, and Pasternak]{loa2001pressure}
Loa,~I.; Adler,~P.; Grzechnik,~A.; Syassen,~K.; Schwarz,~U.; Hanfland,~M.;
  Rozenberg,~G.~K.; Gorodetsky,~P.; Pasternak,~M. Pressure-induced quenching of
  the {J}ahn-{T}eller distortion and insulator-to-metal transition in
  {LaMnO}$_3$. \emph{Phys. Rev. Lett.} \textbf{2001}, \emph{87}, 125501\relax
\mciteBstWouldAddEndPuncttrue
\mciteSetBstMidEndSepPunct{\mcitedefaultmidpunct}
{\mcitedefaultendpunct}{\mcitedefaultseppunct}\relax
\EndOfBibitem
\bibitem[Zhou \latin{et~al.}(2011)Zhou, Alonso, Han, Fern{\'a}ndez-D{\'\i}az,
  Cheng, and Goodenough]{zhou2011jahn}
Zhou,~J.-S.; Alonso,~J.; Han,~J.; Fern{\'a}ndez-D{\'\i}az,~M.; Cheng,~J.-G.;
  Goodenough,~J. Jahn-{T}eller distortion in perovskite {KCuF}$_3$ under high
  pressure. \emph{Journal of Fluorine Chemistry} \textbf{2011}, \emph{132},
  1117--1121\relax
\mciteBstWouldAddEndPuncttrue
\mciteSetBstMidEndSepPunct{\mcitedefaultmidpunct}
{\mcitedefaultendpunct}{\mcitedefaultseppunct}\relax
\EndOfBibitem
\bibitem[Caslin \latin{et~al.}(2016)Caslin, Kremer, Razavi, Hanfland, Syassen,
  Gordon, and Whangbo]{caslin2016competing}
Caslin,~K.; Kremer,~R.; Razavi,~F.; Hanfland,~M.; Syassen,~K.; Gordon,~E.;
  Whangbo,~M.-H. Competing {J}ahn-{T}eller distortions and hydrostatic pressure
  effects in the quasi-one-dimensional quantum ferromagnet {C}u{A}s$_2${O}$_4$.
  \emph{Phys. Rev. B} \textbf{2016}, \emph{93}, 022301\relax
\mciteBstWouldAddEndPuncttrue
\mciteSetBstMidEndSepPunct{\mcitedefaultmidpunct}
{\mcitedefaultendpunct}{\mcitedefaultseppunct}\relax
\EndOfBibitem
\bibitem[Collings \latin{et~al.}(2018)Collings, Bykov, Bykova, Hanfland, van
  Smaalen, Dubrovinsky, and Dubrovinskaia]{collings2018disorder}
Collings,~I.~E.; Bykov,~M.; Bykova,~E.; Hanfland,~M.; van Smaalen,~S.;
  Dubrovinsky,~L.; Dubrovinskaia,~N. Disorder--order transitions in the
  perovskite metal--organic frameworks
  [({CH}$_3$)$_2${NH}$_2$][{M}({HCOO})$_3$] at high pressure.
  \emph{CrystEngComm} \textbf{2018}, \emph{20}, 3512--3521\relax
\mciteBstWouldAddEndPuncttrue
\mciteSetBstMidEndSepPunct{\mcitedefaultmidpunct}
{\mcitedefaultendpunct}{\mcitedefaultseppunct}\relax
\EndOfBibitem
\bibitem[Bostr{\"o}m \latin{et~al.}(2019)Bostr{\"o}m, Collings, Cairns, Romao,
  and Goodwin]{bostrom2019high}
Bostr{\"o}m,~H.~L.; Collings,~I.~E.; Cairns,~A.~B.; Romao,~C.~P.;
  Goodwin,~A.~L. High-pressure behaviour of {P}russian blue analogues:
  interplay of hydration, {J}ahn-{T}eller distortions and vacancies.
  \emph{Dalton Transactions} \textbf{2019}, \emph{48}, 1647--1655\relax
\mciteBstWouldAddEndPuncttrue
\mciteSetBstMidEndSepPunct{\mcitedefaultmidpunct}
{\mcitedefaultendpunct}{\mcitedefaultseppunct}\relax
\EndOfBibitem
\bibitem[Pinsard-Gaudart \latin{et~al.}(2001)Pinsard-Gaudart,
  Rodriguez-Carvajal, Daoud-Aladine, Goncharenko, Medarde, Smith, and
  Revcolevschi]{pinsard2001stability}
Pinsard-Gaudart,~L.; Rodriguez-Carvajal,~J.; Daoud-Aladine,~A.;
  Goncharenko,~I.; Medarde,~M.; Smith,~R.; Revcolevschi,~A. Stability of the
  {J}ahn-{T}eller effect and magnetic study of {LaMnO}$_3$ under pressure.
  \emph{Phys. Rev. B} \textbf{2001}, \emph{64}, 064426\relax
\mciteBstWouldAddEndPuncttrue
\mciteSetBstMidEndSepPunct{\mcitedefaultmidpunct}
{\mcitedefaultendpunct}{\mcitedefaultseppunct}\relax
\EndOfBibitem
\bibitem[Rodriguez-Carvajal \latin{et~al.}(1998)Rodriguez-Carvajal, Hennion,
  Moussa, Moudden, Pinsard, and Revcolevschi]{rodriguez1998neutron}
Rodriguez-Carvajal,~J.; Hennion,~M.; Moussa,~F.; Moudden,~A.; Pinsard,~L.;
  Revcolevschi,~A. Neutron-diffraction study of the {J}ahn-{T}eller transition
  in stoichiometric {LaMnO}$_3$. \emph{Phys. Rev. B} \textbf{1998}, \emph{57},
  R3189\relax
\mciteBstWouldAddEndPuncttrue
\mciteSetBstMidEndSepPunct{\mcitedefaultmidpunct}
{\mcitedefaultendpunct}{\mcitedefaultseppunct}\relax
\EndOfBibitem
\bibitem[Zhou and Goodenough(1999)Zhou, and Goodenough]{zhou1999paramagnetic}
Zhou,~J.-S.; Goodenough,~J. Paramagnetic phase in single-crystal
  {L}a{M}n{O}$_3$. \emph{Phys. Rev. B} \textbf{1999}, \emph{60}, R15002\relax
\mciteBstWouldAddEndPuncttrue
\mciteSetBstMidEndSepPunct{\mcitedefaultmidpunct}
{\mcitedefaultendpunct}{\mcitedefaultseppunct}\relax
\EndOfBibitem
\bibitem[Ovsyannikov \latin{et~al.}(2021)Ovsyannikov, Aslandukova, Aslandukov,
  Chariton, Tsirlin, Korobeynikov, Morozova, Fedotenko, Khandarkhaeva, and
  Dubrovinsky]{ovsyannikov2021structural}
Ovsyannikov,~S.~V.; Aslandukova,~A.~A.; Aslandukov,~A.; Chariton,~S.;
  Tsirlin,~A.~A.; Korobeynikov,~I.~V.; Morozova,~N.~V.; Fedotenko,~T.;
  Khandarkhaeva,~S.; Dubrovinsky,~L. Structural stability and properties of
  marokite-type $\gamma$-{M}n$_3${O}$_4$. \emph{Inorganic Chemistry}
  \textbf{2021}, \emph{60}, 13440--13452\relax
\mciteBstWouldAddEndPuncttrue
\mciteSetBstMidEndSepPunct{\mcitedefaultmidpunct}
{\mcitedefaultendpunct}{\mcitedefaultseppunct}\relax
\EndOfBibitem
\bibitem[Paris \latin{et~al.}(1992)Paris, Ross~Il, and Olijnyk]{paris1992mn3o4}
Paris,~E.; Ross~Il,~C.~R.; Olijnyk,~H. Mn$_{3}${O}$_4$ at high pressure: a
  diamond-anvil cell study and a structural modelling. \emph{European Journal
  of Mineralogy} \textbf{1992}, 87--94\relax
\mciteBstWouldAddEndPuncttrue
\mciteSetBstMidEndSepPunct{\mcitedefaultmidpunct}
{\mcitedefaultendpunct}{\mcitedefaultseppunct}\relax
\EndOfBibitem
\bibitem[Li \latin{et~al.}(2020)Li, Liu, Dong, Li, Dong, Lin, Liu, Wang, Shen,
  Li, \latin{et~al.} others]{li2020size}
Li,~J.; Liu,~B.; Dong,~J.; Li,~C.; Dong,~Q.; Lin,~T.; Liu,~R.; Wang,~P.;
  Shen,~P.; Li,~Q., \latin{et~al.}  Size and morphology effects on the high
  pressure behaviors of {M}n$_3${O}$_4$ nanorods. \emph{Nanoscale Advances}
  \textbf{2020}, \emph{2}, 5841--5847\relax
\mciteBstWouldAddEndPuncttrue
\mciteSetBstMidEndSepPunct{\mcitedefaultmidpunct}
{\mcitedefaultendpunct}{\mcitedefaultseppunct}\relax
\EndOfBibitem
\bibitem[{\AA}sbrink \latin{et~al.}(1999){\AA}sbrink, Wa{\'s}kowska, Gerward,
  Olsen, and Talik]{aasbrink1999high}
{\AA}sbrink,~S.; Wa{\'s}kowska,~A.; Gerward,~L.; Olsen,~J.~S.; Talik,~E.
  High-pressure phase transition and properties of spinel {Z}n{M}n$_2${O}$_4$.
  \emph{Phys. Rev. B} \textbf{1999}, \emph{60}, 12651\relax
\mciteBstWouldAddEndPuncttrue
\mciteSetBstMidEndSepPunct{\mcitedefaultmidpunct}
{\mcitedefaultendpunct}{\mcitedefaultseppunct}\relax
\EndOfBibitem
\bibitem[Choi \latin{et~al.}(2006)Choi, Shim, and Min]{choi2006electronic}
Choi,~H.; Shim,~J.; Min,~B. Electronic structures and magnetic properties of
  spinel {Z}n{M}n$_2${O}$_4$ under high pressure. \emph{Phys. Rev. B}
  \textbf{2006}, \emph{74}, 172103\relax
\mciteBstWouldAddEndPuncttrue
\mciteSetBstMidEndSepPunct{\mcitedefaultmidpunct}
{\mcitedefaultendpunct}{\mcitedefaultseppunct}\relax
\EndOfBibitem
\bibitem[Lawler \latin{et~al.}(2021)Lawler, Smith, Evans, Dos~Santos, Molaison,
  Bos, Mutka, Henry, Argyriou, Salamat, \latin{et~al.}
  others]{lawler2021decoupling}
Lawler,~K.~V.; Smith,~D.; Evans,~S.~R.; Dos~Santos,~A.~M.; Molaison,~J.~J.;
  Bos,~J.-W.~G.; Mutka,~H.; Henry,~P.~F.; Argyriou,~D.~N.; Salamat,~A.,
  \latin{et~al.}  Decoupling Lattice and Magnetic Instabilities in Frustrated
  {C}u{M}n{O}$_2$. \emph{Inorganic Chemistry} \textbf{2021}, \emph{60},
  6004--6015\relax
\mciteBstWouldAddEndPuncttrue
\mciteSetBstMidEndSepPunct{\mcitedefaultmidpunct}
{\mcitedefaultendpunct}{\mcitedefaultseppunct}\relax
\EndOfBibitem
\bibitem[Medarde \latin{et~al.}(1997)Medarde, Mesot, Rosenkranz, Lacorre,
  Marshall, Klotz, Loveday, Hamel, Hull, and Radaelli]{medarde1997pressure}
Medarde,~M.; Mesot,~J.; Rosenkranz,~S.; Lacorre,~P.; Marshall,~W.; Klotz,~S.;
  Loveday,~J.; Hamel,~G.; Hull,~S.; Radaelli,~P. Pressure-induced
  orthorhombic-rhombohedral phase transition in {NdNiO}$_3$. \emph{Physica B:
  Condensed Matter} \textbf{1997}, \emph{234}, 15--17\relax
\mciteBstWouldAddEndPuncttrue
\mciteSetBstMidEndSepPunct{\mcitedefaultmidpunct}
{\mcitedefaultendpunct}{\mcitedefaultseppunct}\relax
\EndOfBibitem
\bibitem[Garc{\'\i}a-Mu{\~n}oz \latin{et~al.}(1994)Garc{\'\i}a-Mu{\~n}oz,
  Rodr{\'\i}guez-Carvajal, and Lacorre]{garcia1994neutron}
Garc{\'\i}a-Mu{\~n}oz,~J.; Rodr{\'\i}guez-Carvajal,~J.; Lacorre,~P.
  Neutron-diffraction study of the magnetic ordering in the insulating regime
  of the perovskites \textit{R}{NiO}$_3$ (\textit{R}= {P}r and {N}d).
  \emph{Phys. Rev. B} \textbf{1994}, \emph{50}, 978\relax
\mciteBstWouldAddEndPuncttrue
\mciteSetBstMidEndSepPunct{\mcitedefaultmidpunct}
{\mcitedefaultendpunct}{\mcitedefaultseppunct}\relax
\EndOfBibitem
\bibitem[Mizokawa \latin{et~al.}(2000)Mizokawa, Khomskii, and
  Sawatzky]{mizokawa2000spin}
Mizokawa,~T.; Khomskii,~D.; Sawatzky,~G. Spin and charge ordering in self-doped
  {M}ott insulators. \emph{Phys. Rev. B} \textbf{2000}, \emph{61}, 11263\relax
\mciteBstWouldAddEndPuncttrue
\mciteSetBstMidEndSepPunct{\mcitedefaultmidpunct}
{\mcitedefaultendpunct}{\mcitedefaultseppunct}\relax
\EndOfBibitem
\bibitem[Garc{\'\i}a-Mu{\~n}oz \latin{et~al.}(2009)Garc{\'\i}a-Mu{\~n}oz,
  Aranda, Alonso, and Mart{\'\i}nez-Lope]{garcia2009structure}
Garc{\'\i}a-Mu{\~n}oz,~J.; Aranda,~M.; Alonso,~J.; Mart{\'\i}nez-Lope,~M.
  Structure and charge order in the antiferromagnetic band-insulating phase of
  {NdNiO}$_3$. \emph{Phys. Rev. B} \textbf{2009}, \emph{79}, 134432\relax
\mciteBstWouldAddEndPuncttrue
\mciteSetBstMidEndSepPunct{\mcitedefaultmidpunct}
{\mcitedefaultendpunct}{\mcitedefaultseppunct}\relax
\EndOfBibitem
\bibitem[Johnston \latin{et~al.}(2014)Johnston, Mukherjee, Elfimov, Berciu, and
  Sawatzky]{johnston2014charge}
Johnston,~S.; Mukherjee,~A.; Elfimov,~I.; Berciu,~M.; Sawatzky,~G.~A. Charge
  disproportionation without charge transfer in the rare-earth-element
  nickelates as a possible mechanism for the metal-insulator transition.
  \emph{Phys. Rev. Lett.} \textbf{2014}, \emph{112}, 106404\relax
\mciteBstWouldAddEndPuncttrue
\mciteSetBstMidEndSepPunct{\mcitedefaultmidpunct}
{\mcitedefaultendpunct}{\mcitedefaultseppunct}\relax
\EndOfBibitem
\bibitem[Wawrzy{\'n}ska \latin{et~al.}(2007)Wawrzy{\'n}ska, Coldea, Wheeler,
  Mazin, Johannes, S{\"o}rgel, Jansen, Ibberson, and
  Radaelli]{wawrzynska2007orbital}
Wawrzy{\'n}ska,~E.; Coldea,~R.; Wheeler,~E.~M.; Mazin,~I.~I.; Johannes,~M.;
  S{\"o}rgel,~T.; Jansen,~M.; Ibberson,~R.~M.; Radaelli,~P.~G. Orbital
  degeneracy removed by charge order in triangular antiferromagnet {AgNiO}$_2$.
  \emph{Phys. Rev. Lett.} \textbf{2007}, \emph{99}, 157204\relax
\mciteBstWouldAddEndPuncttrue
\mciteSetBstMidEndSepPunct{\mcitedefaultmidpunct}
{\mcitedefaultendpunct}{\mcitedefaultseppunct}\relax
\EndOfBibitem
\bibitem[Kang \latin{et~al.}(2007)Kang, Lee, Kim, Lee, Song, Shin, Han, Hwang,
  Jung, Shin, \latin{et~al.} others]{kang2007valence}
Kang,~J.-S.; Lee,~S.; Kim,~G.; Lee,~H.; Song,~H.; Shin,~Y.; Han,~S.; Hwang,~C.;
  Jung,~M.; Shin,~H., \latin{et~al.}  Valence and spin states in delafossite
  {A}g{N}i{O}$_2$ and the frustrated {J}ahn-{T}eller system
  \textit{A}{N}i{O}$_2$ (\textit{A}= {L}i, {N}a). \emph{Phys. Rev. B}
  \textbf{2007}, \emph{76}, 195122\relax
\mciteBstWouldAddEndPuncttrue
\mciteSetBstMidEndSepPunct{\mcitedefaultmidpunct}
{\mcitedefaultendpunct}{\mcitedefaultseppunct}\relax
\EndOfBibitem
\bibitem[Rougier \latin{et~al.}(1995)Rougier, Delmas, and
  Chadwick]{rougier1995non}
Rougier,~A.; Delmas,~C.; Chadwick,~A.~V. Non-cooperative {J}ahn-{T}eller effect
  in {LiNiO}$_2$: an {EXAFS} study. \emph{Solid State Communications}
  \textbf{1995}, \emph{94}, 123--127\relax
\mciteBstWouldAddEndPuncttrue
\mciteSetBstMidEndSepPunct{\mcitedefaultmidpunct}
{\mcitedefaultendpunct}{\mcitedefaultseppunct}\relax
\EndOfBibitem
\bibitem[Chung \latin{et~al.}(2005)Chung, Proffen, Shamoto, Ghorayeb,
  Croguennec, Tian, Sales, Jin, Mandrus, and Egami]{chung2005local}
Chung,~J.-H.; Proffen,~T.; Shamoto,~S.; Ghorayeb,~A.; Croguennec,~L.; Tian,~W.;
  Sales,~B.~C.; Jin,~R.; Mandrus,~D.; Egami,~T. Local structure of {LiNiO}$_2$
  studied by neutron diffraction. \emph{Phys. Rev. B} \textbf{2005}, \emph{71},
  064410\relax
\mciteBstWouldAddEndPuncttrue
\mciteSetBstMidEndSepPunct{\mcitedefaultmidpunct}
{\mcitedefaultendpunct}{\mcitedefaultseppunct}\relax
\EndOfBibitem
\bibitem[Chen \latin{et~al.}(2011)Chen, Freeman, and Harding]{chen2011charge}
Chen,~H.; Freeman,~C.~L.; Harding,~J.~H. Charge disproportionation and
  {J}ahn-{T}eller distortion in {LiNiO}$_2$ and {NaNiO}$_2$: A density
  functional theory study. \emph{Phys. Rev. B} \textbf{2011}, \emph{84},
  085108\relax
\mciteBstWouldAddEndPuncttrue
\mciteSetBstMidEndSepPunct{\mcitedefaultmidpunct}
{\mcitedefaultendpunct}{\mcitedefaultseppunct}\relax
\EndOfBibitem
\bibitem[Foyevtsova \latin{et~al.}(2019)Foyevtsova, Elfimov, Rottler, and
  Sawatzky]{foyevtsova2019linio}
Foyevtsova,~K.; Elfimov,~I.; Rottler,~J.; Sawatzky,~G.~A. {LiNiO}$_2$ as a
  high-entropy charge-and bond-disproportionated glass. \emph{Phys. Rev. B}
  \textbf{2019}, \emph{100}, 165104\relax
\mciteBstWouldAddEndPuncttrue
\mciteSetBstMidEndSepPunct{\mcitedefaultmidpunct}
{\mcitedefaultendpunct}{\mcitedefaultseppunct}\relax
\EndOfBibitem
\bibitem[Green \latin{et~al.}(2020)Green, Wadati, Regier, Achkar, McMahon,
  Clancy, Dabkowska, Gaulin, Sawatzky, and Hawthorn]{green2020evidence}
Green,~R.; Wadati,~H.; Regier,~T.; Achkar,~A.; McMahon,~C.; Clancy,~J.;
  Dabkowska,~H.; Gaulin,~B.; Sawatzky,~G.; Hawthorn,~D. Evidence for
  bond-disproportionation in {LiNiO}$_2$ from x-ray absorption spectroscopy.
  \emph{arXiv preprint arXiv:2011.06441} \textbf{2020}, \relax
\mciteBstWouldAddEndPunctfalse
\mciteSetBstMidEndSepPunct{\mcitedefaultmidpunct}
{}{\mcitedefaultseppunct}\relax
\EndOfBibitem
\bibitem[Zhou and Goodenough(2004)Zhou, and Goodenough]{zhou2004chemical}
Zhou,~J.-S.; Goodenough,~J. Chemical bonding and electronic structure of
  \textit{R}{NiO}$_3$ (\textit{R}= rare earth). \emph{Phys. Rev. B}
  \textbf{2004}, \emph{69}, 153105\relax
\mciteBstWouldAddEndPuncttrue
\mciteSetBstMidEndSepPunct{\mcitedefaultmidpunct}
{\mcitedefaultendpunct}{\mcitedefaultseppunct}\relax
\EndOfBibitem
\bibitem[Dick \latin{et~al.}(1997)Dick, M{\"u}ller, Preissinger, and
  Zeiske]{dick1997structure}
Dick,~S.; M{\"u}ller,~M.; Preissinger,~F.; Zeiske,~T. The structure of
  monoclinic {NaNiO}$_2$ as determined by powder X-ray and neutron scattering.
  \emph{Powder Diffraction} \textbf{1997}, \emph{12}, 239--241\relax
\mciteBstWouldAddEndPuncttrue
\mciteSetBstMidEndSepPunct{\mcitedefaultmidpunct}
{\mcitedefaultendpunct}{\mcitedefaultseppunct}\relax
\EndOfBibitem
\bibitem[Chappel \latin{et~al.}(2000)Chappel, Nunez-Regueiro, Chouteau, Isnard,
  and Darie]{chappel2000study}
Chappel,~E.; Nunez-Regueiro,~M.; Chouteau,~G.; Isnard,~O.; Darie,~C. Study of
  the ferrodistorsive orbital ordering in {NaNiO}$_2$ by neutron diffraction
  and submillimeter wave {ESR}. \emph{The Euro. Phys. Journ. B-Cond. Matt. and
  Complex Systems} \textbf{2000}, \emph{17}, 615--622\relax
\mciteBstWouldAddEndPuncttrue
\mciteSetBstMidEndSepPunct{\mcitedefaultmidpunct}
{\mcitedefaultendpunct}{\mcitedefaultseppunct}\relax
\EndOfBibitem
\bibitem[Galakhov \latin{et~al.}(1995)Galakhov, Kurmaev, Neumann, Kellerman,
  Gorshkov, \latin{et~al.} others]{galakhov1995electronic}
Galakhov,~V.; Kurmaev,~E.; Neumann,~M.; Kellerman,~D.; Gorshkov,~V.,
  \latin{et~al.}  Electronic structure of {LiNiO}$_2$, {LiFeO}$_2$ and
  {LiCrO}$_2$: {X}-ray photoelectron and X-ray emission study. \emph{Solid
  State Communications} \textbf{1995}, \emph{95}, 347--351\relax
\mciteBstWouldAddEndPuncttrue
\mciteSetBstMidEndSepPunct{\mcitedefaultmidpunct}
{\mcitedefaultendpunct}{\mcitedefaultseppunct}\relax
\EndOfBibitem
\bibitem[Lewis \latin{et~al.}(2005)Lewis, Gaulin, Filion, Kallin, Berlinsky,
  Dabkowska, Qiu, and Copley]{lewis2005ordering}
Lewis,~M.; Gaulin,~B.; Filion,~L.; Kallin,~C.; Berlinsky,~A.; Dabkowska,~H.;
  Qiu,~Y.; Copley,~J. Ordering and spin waves in {NaNiO}$_2$: A stacked quantum
  ferromagnet. \emph{Phys. Rev. B} \textbf{2005}, \emph{72}, 014408\relax
\mciteBstWouldAddEndPuncttrue
\mciteSetBstMidEndSepPunct{\mcitedefaultmidpunct}
{\mcitedefaultendpunct}{\mcitedefaultseppunct}\relax
\EndOfBibitem
\bibitem[Baker \latin{et~al.}(2005)Baker, Lancaster, Blundell, Brooks, Hayes,
  Prabhakaran, and Pratt]{baker2005thermodynamic}
Baker,~P.; Lancaster,~T.; Blundell,~S.; Brooks,~M.; Hayes,~W.; Prabhakaran,~D.;
  Pratt,~F. Thermodynamic and magnetic properties of the layered triangular
  magnet {NaNiO}$_2$. \emph{Phys. Rev. B} \textbf{2005}, \emph{72},
  104414\relax
\mciteBstWouldAddEndPuncttrue
\mciteSetBstMidEndSepPunct{\mcitedefaultmidpunct}
{\mcitedefaultendpunct}{\mcitedefaultseppunct}\relax
\EndOfBibitem
\bibitem[Darie \latin{et~al.}(2005)Darie, Bordet, De~Brion, Holzapfel, Isnard,
  Lecchi, Lorenzo, and Suard]{darie2005magnetic}
Darie,~C.; Bordet,~P.; De~Brion,~S.; Holzapfel,~M.; Isnard,~O.; Lecchi,~A.;
  Lorenzo,~J.; Suard,~E. Magnetic structure of the spin-1/2 layer compound
  {NaNiO}$_2$. \emph{The Euro. Phys. Journ. B-Cond. Matt. and Complex Systems}
  \textbf{2005}, \emph{43}, 159--162\relax
\mciteBstWouldAddEndPuncttrue
\mciteSetBstMidEndSepPunct{\mcitedefaultmidpunct}
{\mcitedefaultendpunct}{\mcitedefaultseppunct}\relax
\EndOfBibitem
\bibitem[Vassilaras \latin{et~al.}(2012)Vassilaras, Ma, Li, and
  Ceder]{vassilaras2012electrochemical}
Vassilaras,~P.; Ma,~X.; Li,~X.; Ceder,~G. Electrochemical properties of
  monoclinic {NaNiO}$_2$. \emph{Journal of The Electrochem. Soc.}
  \textbf{2012}, \emph{160}, A207\relax
\mciteBstWouldAddEndPuncttrue
\mciteSetBstMidEndSepPunct{\mcitedefaultmidpunct}
{\mcitedefaultendpunct}{\mcitedefaultseppunct}\relax
\EndOfBibitem
\bibitem[Han \latin{et~al.}(2014)Han, Gonzalo, Casas-Cabanas, and
  Rojo]{han2014structural}
Han,~M.~H.; Gonzalo,~E.; Casas-Cabanas,~M.; Rojo,~T. Structural evolution and
  electrochemistry of monoclinic {NaNiO}$_2$ upon the first cycling process.
  \emph{Journal of Power Sources} \textbf{2014}, \emph{258}, 266--271\relax
\mciteBstWouldAddEndPuncttrue
\mciteSetBstMidEndSepPunct{\mcitedefaultmidpunct}
{\mcitedefaultendpunct}{\mcitedefaultseppunct}\relax
\EndOfBibitem
\bibitem[Neuefeind \latin{et~al.}(2012)Neuefeind, Feygenson, Carruth, Hoffmann,
  and Chipley]{neuefeind2012nanoscale}
Neuefeind,~J.; Feygenson,~M.; Carruth,~J.; Hoffmann,~R.; Chipley,~K.~K. The
  nanoscale ordered materials diffractometer {NOMAD} at the spallation neutron
  source {SNS}. \emph{Nuclear Instruments and Methods in Physics Research
  Section B: Beam Interactions with Materials and Atoms} \textbf{2012},
  \emph{287}, 68--75\relax
\mciteBstWouldAddEndPuncttrue
\mciteSetBstMidEndSepPunct{\mcitedefaultmidpunct}
{\mcitedefaultendpunct}{\mcitedefaultseppunct}\relax
\EndOfBibitem
\bibitem[Bull \latin{et~al.}(2016)Bull, Funnell, Tucker, Hull, Francis, and
  Marshall]{bull2016pearl}
Bull,~C.~L.; Funnell,~N.~P.; Tucker,~M.~G.; Hull,~S.; Francis,~D.~J.;
  Marshall,~W.~G. P{EARL}: the high pressure neutron powder diffractometer at
  {ISIS}. \emph{High Pressure Research} \textbf{2016}, \emph{36},
  493--511\relax
\mciteBstWouldAddEndPuncttrue
\mciteSetBstMidEndSepPunct{\mcitedefaultmidpunct}
{\mcitedefaultendpunct}{\mcitedefaultseppunct}\relax
\EndOfBibitem
\bibitem[Fortes(2019)]{fortes_manual_paper}
Fortes,~A.~D. R{AL} Technical Reports, {RAL}-{TR}-2019-002. \textbf{2019},
  \relax
\mciteBstWouldAddEndPunctfalse
\mciteSetBstMidEndSepPunct{\mcitedefaultmidpunct}
{}{\mcitedefaultseppunct}\relax
\EndOfBibitem
\bibitem[Arnold \latin{et~al.}(2014)Arnold, Bilheux, Borreguero, Buts,
  Campbell, Chapon, Doucet, Draper, Leal, Gigg, \latin{et~al.}
  others]{arnold2014mantid}
Arnold,~O.; Bilheux,~J.-C.; Borreguero,~J.; Buts,~A.; Campbell,~S.~I.;
  Chapon,~L.; Doucet,~M.; Draper,~N.; Leal,~R.~F.; Gigg,~M., \latin{et~al.}
  Mantid—{D}ata analysis and visualization package for neutron scattering and
  $\mu$ {SR} experiments. \emph{Nuclear Instruments and Methods in Physics
  Research Section A: Accelerators, Spectrometers, Detectors and Associated
  Equipment} \textbf{2014}, \emph{764}, 156--166\relax
\mciteBstWouldAddEndPuncttrue
\mciteSetBstMidEndSepPunct{\mcitedefaultmidpunct}
{\mcitedefaultendpunct}{\mcitedefaultseppunct}\relax
\EndOfBibitem
\bibitem[Coelho(2018)]{coelho2018topas}
Coelho,~A.~A. {TOPAS} and {TOPAS}-Academic: an optimization program integrating
  computer algebra and crystallographic objects written in {C}++. \emph{Journal
  of Applied Crystallography} \textbf{2018}, \emph{51}, 210--218\relax
\mciteBstWouldAddEndPuncttrue
\mciteSetBstMidEndSepPunct{\mcitedefaultmidpunct}
{\mcitedefaultendpunct}{\mcitedefaultseppunct}\relax
\EndOfBibitem
\bibitem[Pawley(1981)]{pawley1981unit}
Pawley,~G. Unit-cell refinement from powder diffraction scans. \emph{Journal of
  Applied Crystallography} \textbf{1981}, \emph{14}, 357--361\relax
\mciteBstWouldAddEndPuncttrue
\mciteSetBstMidEndSepPunct{\mcitedefaultmidpunct}
{\mcitedefaultendpunct}{\mcitedefaultseppunct}\relax
\EndOfBibitem
\bibitem[Rietveld(1969)]{rietveld1969profile}
Rietveld,~H. A profile refinement method for nuclear and magnetic structures.
  \emph{Journal of Applied Crystallography} \textbf{1969}, \emph{2},
  65--71\relax
\mciteBstWouldAddEndPuncttrue
\mciteSetBstMidEndSepPunct{\mcitedefaultmidpunct}
{\mcitedefaultendpunct}{\mcitedefaultseppunct}\relax
\EndOfBibitem
\bibitem[Thompson \latin{et~al.}(1987)Thompson, Cox, and
  Hastings]{thompson1987rietveld}
Thompson,~P.; Cox,~D.; Hastings,~J. Rietveld refinement of {D}ebye-{S}cherrer
  synchrotron {X}-ray data from {A}l$_2${O}$_3$. \emph{Journal of Applied
  Crystallography} \textbf{1987}, \emph{20}, 79--83\relax
\mciteBstWouldAddEndPuncttrue
\mciteSetBstMidEndSepPunct{\mcitedefaultmidpunct}
{\mcitedefaultendpunct}{\mcitedefaultseppunct}\relax
\EndOfBibitem
\bibitem[Sofin and Jansen(2005)Sofin, and Jansen]{sofin2005new}
Sofin,~M.; Jansen,~M. New route of preparation and properties of {NaNiO}$_2$.
  \emph{Zeitschrift f{\"u}r Naturforschung B} \textbf{2005}, \emph{60},
  701--704\relax
\mciteBstWouldAddEndPuncttrue
\mciteSetBstMidEndSepPunct{\mcitedefaultmidpunct}
{\mcitedefaultendpunct}{\mcitedefaultseppunct}\relax
\EndOfBibitem
\bibitem[Hoppe(1979)]{hoppe1979effective}
Hoppe,~R. Effective coordination numbers ({EC}o{N}) and mean fictive ionic
  radii ({MEFIR}). \emph{Zeitschrift f{\"u}r Kristallographie-Crystalline
  Materials} \textbf{1979}, \emph{150}, 23--52\relax
\mciteBstWouldAddEndPuncttrue
\mciteSetBstMidEndSepPunct{\mcitedefaultmidpunct}
{\mcitedefaultendpunct}{\mcitedefaultseppunct}\relax
\EndOfBibitem
\bibitem[Baur(1974)]{baur1974geometry}
Baur,~W. The geometry of polyhedral distortions. {P}redictive relationships for
  the phosphate group. \emph{Acta Crystallographica Section B: Structural
  Crystallography and Crystal Chemistry} \textbf{1974}, \emph{30},
  1195--1215\relax
\mciteBstWouldAddEndPuncttrue
\mciteSetBstMidEndSepPunct{\mcitedefaultmidpunct}
{\mcitedefaultendpunct}{\mcitedefaultseppunct}\relax
\EndOfBibitem
\bibitem[Kimber(2012)]{kimber2012charge}
Kimber,~S.~A. Charge and orbital order in frustrated
  {P}b$_{3}${M}n$_7${O}$_{15}$. \emph{Journal of Physics: Condensed Matter}
  \textbf{2012}, \emph{24}, 186002\relax
\mciteBstWouldAddEndPuncttrue
\mciteSetBstMidEndSepPunct{\mcitedefaultmidpunct}
{\mcitedefaultendpunct}{\mcitedefaultseppunct}\relax
\EndOfBibitem
\bibitem[Hermann \latin{et~al.}(2017)Hermann, Ebad-Allah, Freund, Pietsch,
  Jesche, Tsirlin, Deisenhofer, Hanfland, Gegenwart, and
  Kuntscher]{hermann2017high}
Hermann,~V.; Ebad-Allah,~J.; Freund,~F.; Pietsch,~I.; Jesche,~A.;
  Tsirlin,~A.~A.; Deisenhofer,~J.; Hanfland,~M.; Gegenwart,~P.;
  Kuntscher,~C.~A. High-pressure versus isoelectronic doping effect on the
  honeycomb iridate {N}a$_2${I}r{O}$_3$. \emph{Phys. Rev. B} \textbf{2017},
  \emph{96}, 195137\relax
\mciteBstWouldAddEndPuncttrue
\mciteSetBstMidEndSepPunct{\mcitedefaultmidpunct}
{\mcitedefaultendpunct}{\mcitedefaultseppunct}\relax
\EndOfBibitem
\bibitem[Layek \latin{et~al.}(2020)Layek, Mehlawat, Levy, Greenberg, Pasternak,
  Iti{\'e}, Singh, and Rozenberg]{layek2020electronic}
Layek,~S.; Mehlawat,~K.; Levy,~D.; Greenberg,~E.; Pasternak,~M.;
  Iti{\'e},~J.-P.; Singh,~Y.; Rozenberg,~G.~K. Electronic and structural
  properties of the honeycomb iridates \textit{A}$_2${I}r{O}$_3$ (\textit{A}=
  {N}a, {L}i) at elevated pressures. \emph{Phys. Rev. B} \textbf{2020},
  \emph{102}, 085156\relax
\mciteBstWouldAddEndPuncttrue
\mciteSetBstMidEndSepPunct{\mcitedefaultmidpunct}
{\mcitedefaultendpunct}{\mcitedefaultseppunct}\relax
\EndOfBibitem
\bibitem[Cliffe and Goodwin(2012)Cliffe, and Goodwin]{cliffe2012pascal}
Cliffe,~M.~J.; Goodwin,~A.~L. P{ASC}al: a principal axis strain calculator for
  thermal expansion and compressibility determination. \emph{Journal of Applied
  Crystallography} \textbf{2012}, \emph{45}, 1321--1329\relax
\mciteBstWouldAddEndPuncttrue
\mciteSetBstMidEndSepPunct{\mcitedefaultmidpunct}
{\mcitedefaultendpunct}{\mcitedefaultseppunct}\relax
\EndOfBibitem
\bibitem[Birch(1947)]{birch1947finite}
Birch,~F. Finite elastic strain of cubic crystals. \emph{Physical Review}
  \textbf{1947}, \emph{71}, 809\relax
\mciteBstWouldAddEndPuncttrue
\mciteSetBstMidEndSepPunct{\mcitedefaultmidpunct}
{\mcitedefaultendpunct}{\mcitedefaultseppunct}\relax
\EndOfBibitem
\bibitem[Zhao \latin{et~al.}(2004)Zhao, Ross, and Angel]{zhao2004new}
Zhao,~J.; Ross,~N.~L.; Angel,~R.~J. New view of the high-pressure behaviour of
  GdFeO$_3$-type perovskites. \emph{Acta Crystallographica Section B:
  Structural Science} \textbf{2004}, \emph{60}, 263--271\relax
\mciteBstWouldAddEndPuncttrue
\mciteSetBstMidEndSepPunct{\mcitedefaultmidpunct}
{\mcitedefaultendpunct}{\mcitedefaultseppunct}\relax
\EndOfBibitem
\bibitem[Angel \latin{et~al.}(2005)Angel, Zhao, and Ross]{angel2005general}
Angel,~R.~J.; Zhao,~J.; Ross,~N.~L. General rules for predicting phase
  transitions in perovskites due to octahedral tilting. \emph{Phys. Rev. Lett.}
  \textbf{2005}, \emph{95}, 025503\relax
\mciteBstWouldAddEndPuncttrue
\mciteSetBstMidEndSepPunct{\mcitedefaultmidpunct}
{\mcitedefaultendpunct}{\mcitedefaultseppunct}\relax
\EndOfBibitem
\bibitem[Chen \latin{et~al.}(2011)Chen, Zou, Zhu, Zou, and Cao]{chen2011first}
Chen,~Z.; Zou,~H.; Zhu,~X.; Zou,~J.; Cao,~J. First-principle investigation of
  {J}ahn-{T}eller distortion and topological analysis of chemical bonds in
  {LiNiO}$_2$. \emph{Journal of Solid State Chemistry} \textbf{2011},
  \emph{184}, 1784--1790\relax
\mciteBstWouldAddEndPuncttrue
\mciteSetBstMidEndSepPunct{\mcitedefaultmidpunct}
{\mcitedefaultendpunct}{\mcitedefaultseppunct}\relax
\EndOfBibitem
\bibitem[Ortega-San-Martin \latin{et~al.}(2007)Ortega-San-Martin, Williams,
  Rodgers, Attfield, Heymann, and Huppertz]{ortega2007microstrain}
Ortega-San-Martin,~L.; Williams,~A.~J.; Rodgers,~J.; Attfield,~J.~P.;
  Heymann,~G.; Huppertz,~H. Microstrain sensitivity of orbital and electronic
  phase separation in {SrCrO}$_3$. \emph{Phys. Rev. Lett.} \textbf{2007},
  \emph{99}, 255701\relax
\mciteBstWouldAddEndPuncttrue
\mciteSetBstMidEndSepPunct{\mcitedefaultmidpunct}
{\mcitedefaultendpunct}{\mcitedefaultseppunct}\relax
\EndOfBibitem
\bibitem[Eto \latin{et~al.}(2000)Eto, Endo, Imai, Katayama, and
  Kikegawa]{eto2000crystal}
Eto,~T.; Endo,~S.; Imai,~M.; Katayama,~Y.; Kikegawa,~T. Crystal structure of
  {N}i{O} under high pressure. \emph{Phys. Rev. B} \textbf{2000}, \emph{61},
  14984\relax
\mciteBstWouldAddEndPuncttrue
\mciteSetBstMidEndSepPunct{\mcitedefaultmidpunct}
{\mcitedefaultendpunct}{\mcitedefaultseppunct}\relax
\EndOfBibitem
\bibitem[htt(2021)]{https://doi.org/10.5286/isis.e.rb2000219}
L. {A}. {V}. {N}agle-{C}occo, {S}. {E}. {D}utton, and {C}. {L}. {B}ull.
  \textit{DOI:10.5286/ISIS.E.RB2000219}. {B}ehaviour of {J}ahn-{T}eller
  distorted {NiO}$_6$ octahedra in {NaNiO}$_2$ under pressure. 2021;
  \url{https://data.isis.stfc.ac.uk/doi/STUDY/113601628/}\relax
\mciteBstWouldAddEndPuncttrue
\mciteSetBstMidEndSepPunct{\mcitedefaultmidpunct}
{\mcitedefaultendpunct}{\mcitedefaultseppunct}\relax
\EndOfBibitem
\bibitem[htt(2022)]{https://doi.org/10.17863/CAM.81605}
L. {A}. {V}. {N}agle-{C}occo, {C}. {L}. {B}ull, {C}hristopher {J}. {R}idley,
  {S}. {E}. {D}utton. \textit{DOI:10.17863/CAM.81605}. {D}ata associated with
  "Pressure tuning the Jahn-Teller transition temperature in {NaN}iO$_2$".
  2022; \url{doi.org/10.17863/CAM.81605}\relax
\mciteBstWouldAddEndPuncttrue
\mciteSetBstMidEndSepPunct{\mcitedefaultmidpunct}
{\mcitedefaultendpunct}{\mcitedefaultseppunct}\relax
\EndOfBibitem
\end{mcitethebibliography}

\end{document}